\documentclass[prd,10pt,twocolumn,aps,superscriptaddress,nofootinbib,notitlepage,floatfix,amssymb,amsmath,preprintnumbers]{revtex4}
\usepackage{epsfig}

\newcommand{\beqn}{\begin{eqnarray}}
\newcommand{\eeqn}{\end{eqnarray}}

\newcommand{\beqs}{\begin{subequations}}
\newcommand{\eeqs}{\end{subequations}\\[-2mm]\noindent}

\newcommand{\eq}[1]{(\ref{#1})}

\newcommand{\bs}{\boldsymbol}

\newcommand{\e}{\,{\bs {\mathrm{e}}}}
\newcommand{\nab}{{\bs \nabla}}

\newcommand{\Z}{{\mathbb Z}}
\newcommand{\R}{{\mathbb R}}
\newcommand{\N}{{\mathbb N}}
\newcommand{\bc}{{\mathfrak{b}}}

\newcommand{\TM}{{\mathrm{TM}}}
\newcommand{\TE}{{\mathrm{TE}}}

\newcommand{\cA}{{\mathcal A}}
\newcommand{\cB}{{\mathcal B}}

\newcommand{\cE}{{\mathcal E}}

\newcommand{\cO}{{\mathcal O}}
\newcommand{\cG}{{\mathcal G}}
\newcommand{\cH}{{\mathcal H}}

\newcommand{\bcA}{{\bs {\mathcal A}}}

\newcommand{\bl}{$\blacktriangleright$}

\newcommand{\avr}[1]{{\langle #1 \rangle}}

\newcommand{\vect}[3]{\left(\begin{array}{c} #1 \\ #2 \\ #3 \end{array} \right) }
\newcommand{\vectt}[2]{\left(\begin{array}{c} #1 \\ #2 \end{array} \right) }
\newcommand{\ket}[1]{{| #1 \rangle}}
\newcommand{\bra}[1]{{\langle #1 |}}

\begin{document}

\title{The Zilch Vortical Effect}
\author{M.N. Chernodub}
\affiliation{Institut Denis Poisson UMR 7013, Universit\'e de Tours, 37200 France}
\affiliation{Laboratory of Physics of Living Matter, Far Eastern Federal University, Sukhanova 8, Vladivostok, 690950, Russia}
\author{Alberto Cortijo}
\affiliation{Materials Science Factory, Instituto de Ciencia de Materiales de Madrid, CSIC, Cantoblanco, 28049 Madrid, Spain}
\author{Karl Landsteiner}
\affiliation{Instituto de F\'{\i}sica Te\'orica UAM/CSIC, C/ Nicol\'as Cabrera 13-15, Cantoblanco, 28049 Madrid, Spain}

\date{July 27, 2018}

\begin{abstract}
We study the question if a helicity transporting current is generated in a rotating photon gas at finite temperature. One problem is that there is no gauge invariant local notion of helicity or helicity current. We circumvent this by studying not only the optical helicity current but also the gauge-invariant ``zilch'' current.  In order to avoid problems of causality, we quantize the system on a cylinder of a finite radius and then discuss the limit of infinite radius. We find that net helicity- and zilch currents are only generated in the case of the finite radius and are due to duality violating boundary conditions. A universal result exists for the current density on the axes of rotation in the high-temperature limit. To lowest order in the angular velocity, it takes a form similar to the well-known temperature dependence of the chiral vortical effect for chiral fermions. We briefly discuss possible relations to gravitational anomalies. 
\end{abstract}
\preprint{IFT-UAM/CSIC-18-085}
\maketitle

\section{Introduction}

The quantum field theory of chiral fermions predicts a number of exotic transport phenomena such as the generation of a current in magnetic field or under rotation. These are known as chiral magnetic and chiral vortical effects (see \cite{Kharzeev:2013ffa,Landsteiner:2016led} for recent reviews). Both effects are related to the presence of chiral anomalies. In particular, the chiral vortical effect is present at finite temperature and can be understood 
{as a signal of} (possibly global) mixed gravitational anomalies \cite{Landsteiner:2011cp, Landsteiner:2011iq,  Jensen:2013rga, Jensen:2012kj, Stone:2018zel, Golkar:2015oxw, Glorioso:2017lcn}.

From the outset it should be emphasized that in a relativistic theory rotation can not be implemented by simply introducing a constant angular velocity in a thermal ensemble \cite{Vilenkin:1980zv}. In infinite space there appears necessarily a region in which the tangential velocity would exceed the speed of light. There are two remedies to this. In a hydrodynamic setup one can consider either localized vortices in the fluid or, {alternatively}, one can restrict the ensemble to a finite space-time region in which no superluminal velocities arise. In infinite space the CVE can be computed by studying an ensemble of rotating fermions and concentrating on the region at the center of rotation~\cite{Vilenkin:1980zv}. Alternatively one can study an ensemble confined with the boundaries of a rotating cylinder 
\cite{Ambrus:2014uqa,Ambrus:2015lfr, Chernodub:2016kxh, Chernodub:2017ref, Chernodub:2017mvp, Ebihara:2016fwa}.

Recently, the question has arisen if and how a similar effect for ensembles of rotating photons could be {made} possible~\cite{Avkhadiev:2017fxj,Yamamoto:2017uul,Zyuzin}. In part, this question can be motivated by the relation of the CVE to the mixed gravitational anomalies as well as by the interesting results on the existence of a similar anomaly for photons~\cite{Dolgov:1988qx, Agullo:2016lkj}. {From a first sight, the case of photons the notion of chirality could naturally be replaced by the concept of helicity}. It turns out, however, that the definition of a {photonic} helicity current analogous to a {fermionic} chiral (or axial) current is much more subtle. A standard way of defining a photonic helicity current is the so--called magnetic helicity, {which in covariant notations can be written as}:
\begin{equation}
J^\mu_{mh} = \epsilon^{\mu\nu\rho\lambda}A_\nu F_{\rho\lambda}\,.
\label{eq:J:mh}
\end{equation}
The drawback of this way of defining helicity is that the current~\eq{eq:J:mh} is neither conserved (since $\partial_\mu J^\mu_{mh} = \tilde F^{\mu\nu}F_{\mu\nu}$ with $\tilde F^{\mu\nu} = \frac{1}{2} \epsilon^{\mu\nu\rho\lambda} F_{\rho\lambda}$) nor gauge invariant. The first inconvenience can be remedied  by defining also the so-called ``optical helicity''~\cite{ref:Calkin}
\begin{equation}
J^\mu_h = \frac 1 2 \epsilon^{\mu\nu\rho\lambda}( A_\nu F_{\rho\lambda} - C_\nu \tilde F_{\rho\lambda})\,,
\label{eq:optical:helicity}
\end{equation}
where $C_\mu$ is a dual gauge potential defined via the relation $\tilde F_{\mu\nu} = \partial_\mu C_\nu - \partial_\nu C_\mu$.
This current is conserved, $\partial_\mu J^\mu_h = 0$, but now there is a new inconvenience: the original $A_\mu$ and dual $C_\mu$ gauge fields are not locally related to each other. As long as one does not insist in a Lorentz invariant Lagrangian formulation of Maxwells equations that might not be considered as a fundamental problem. However the optical helicity is still not gauge invariant and now there are even two gauge symmetries since $C_\mu' = C_\mu + \partial_\mu \theta$ and $C_\mu$ are physically equivalent dual gauge potentials. A gauge invariant global helicity charge $  \mathcal{Q} = \int d^3x J^0_h$ can {still} be defined, but {there is no covariant expression for the helicity density which could be local in terms of original $A_\mu$ and dual $C_\mu$ potential and gauge invariant with respect to both original and dual gauge transformations.}

Fortunately, there are other candidates for physically meaningful measures fo helicity. Quite some time ago Lipkin pointed out that free Maxwell theory allows for an additional conserved quantity~\cite{Lipkin} and soon afterwards Kibble noticed that due to the its nature of a non-interacting theory there is, in fact, an infinite number of such conserved charges~\cite{Kibble}. Following the nomenclature introduced by Lipkin these charges are known as "zilches". The optical helicity can be understood as a particular, non-gauge invariant version of one of these zilch charges. It has also been identified as the generator of electric-magnetic duality transformation~\cite{Deser:1976iy}.

The zilches are (classically) conserved quantities which, except for the optical helicity, have unusual dimensions. We will consider here only the original zilch introduced by Lipkin, a conserved current of dimension five.  While a physical interpretation of the zilch remained obscure for a long time, recently it was shown that the zilch measures the asymmetry in the interaction of the electromagnetic field with small chiral molecules~\cite{tangcohen} {similarly to the effects of the optical helicity on chiral and magnetoelectric media~\cite{Bliokh:2014pva,Bliokh:2018ehc} and Weyl semimetals~\cite{Elbistan:2018fkr}}. We therefore take the zilch as a legitimate local measure {of the helicity of light}.

In order to study the possible realization of a version of the chiral vortical effect for photons we will quantize Maxwell theory on a finite cylinder of radius $R$ and consider an ensemble with a finite fixed angular velocity such $|\Omega R| < 1$. We calculate the thermal averages for the optical helicity current and zilch current along the direction of rotation and study the infinite space limit $R \rightarrow \infty$. It turns out that this infinite space limit is -- in contrast to the fermionic case -- ill defined even if one concentrates on the current at the center of rotation. For finite radius and $|\Omega R|<1$ the ensemble is well defined but the appearance of a non-vanishing total current depends very sensitively on the boundary conditions. 

We will study three types of boundary conditions: perfect electric conducting boundary, perfect magnetic conducting boundary and duality invariant unbounded space. Our finding is that the integrated helicity and zilch currents vanish exactly in the duality invariant case whereas only one type of polarization leads to a non-vanishing net current in the other two cases. More precisely, the Dirichlet boundary conditions on the photon wave functions lead to exactly vanishing net current and only Neumann boundary conditions give rise to a net current. The perfect conducting and dual conducting boundary conditions break however the electric--magnetic duality and therefore introduce a source of helicity or zilch on the boundary. We interpret the net current therefore not as an intrinsic chiral vortical effect but as a result of the symmetry breaking boundary conditions.

This work is organized as follows. In the next section we introduce our notation, the (non-Lorentz covariant) versions of helicity and zilch and associated currents. Then in Section 2 we quantize the Maxwell field in the Coulomb gauge on a cylinder of radius $R$. In Section 3 we study the helicity, zilch and energy currents. We show that the net currents integrated over a cross section of the cylinder vanish for the Dirichlet boundary conditions {on radial functions of photons}. We evaluate numerically the thermal current distributions for different temperatures and angular velocities. Finally, we study the infinite space limit and show that in this limit the current at the axis of rotation is a mathematically ill defined quantity. Indeed, if one tries to evaluate it by an analytic continuation (using inversion relations for polylogarithms) then one finds a complex result. While a truncation to lowest order in angular velocity does give a finite expression it does not coincide with the results previously reported in the literature for the photonic CVE. 

In any case, the physical significance of such a finite result for the photonic CVE is questionable since the resulting integrals for the thermal averages are mathematically well defined only in the strict case $|\Omega| + \epsilon = 1/R$ with $\epsilon >0$. This requirement means that the zero-rotation limit, $\Omega \to 0$, should precede the infinite-volume limit $R \to \infty$. We emphasize that this {requirement sets a stronger constraint for the rotating photons as compared to the case for fermions, because} it arises from the essential property of  bosons that the Bose-Einstein distribution function can take negative values for effectively negative energies, signaling a possible instability towards condensation of low-energy modes. Despite of the difficulties with the calculation in the unbounded domain we find numerically that the high temperature limit of the central current density in the bounded domain does converge to the result in the unbounded domain at linear order in the angular velocity.

We present our conclusions in Section 4. Some of our conventions for vector analysis and useful properties of the Bessel functions are given in the Appendix.  

\section{Photons in nonrotating cylinder}

We consider thermalized photon gas at a fixed temperature $T$ in an infinitely long straight cylindrical volume (often called in the literature as a ``waveguide''). The cylinder has a fixed finite radius $R$ and may rotate around its symmetry axis with the constant angular velocity $\Omega$. For the sake of simplicity we work in the vacuum with permittivity $\varepsilon = 1$ and permeability $\mu = 1$. We also set the speed of light and the reduced Planck constant to unity, $c = \hbar = 1$.

\subsection{System of equations}

\subsubsection{Maxwell equations}

The electromagnetic fields are described by the Maxwell's equations,
\beqs
\beqn
\nab \cdot {\bs E} & = & 0 \,, \\
\nab \cdot {\bs B} & = & 0 \,, \\ 
\nab \times {\bs B} - \frac{\partial {\bs E}}{\partial t} & = & 0\,,\\
\nab \times {\bs E} + \frac{\partial {\bs B}}{\partial t} & = & 0\,,
\eeqn
\label{eq:Maxwell}
\eeqs
where the magnetic field ${\bs B}$ and the electric field ${\bs E}$ are related to the gauge potential $A^\mu = (\phi, {\bs A})$ as follows:
\beqn
{\bs B} = \nab \times {\bs A}\,, \qquad {\bs E} = - \nab \phi - \frac{\partial {\bs A}}{\partial t}\,.
\label{eq:EB:A}
\eeqn

To solve these equations inside a cylinder it is natural to introduce cylindrical coordinates with the radius $\rho$, the azimuthal angle $\varphi$, the height $z$, and the time coordinate $t$ (in the laboratory reference frame). Certain useful formulas of the vector calculus in the cylindrical system of coordinates are summarized in Appendix~\ref{appendix:cylindrical}.

Given the geometry of the problem and the linearity of the Maxwell equations the solutions can be described in the complexified form:
\beqn
{\bs G}(\rho,\varphi,z,t) = e^{- i \omega t + i m \varphi + i k_z z} {\bs G}(\rho)\,,
\label{eq:decomposition:G}
\eeqn
where ${\bs G} = {\bs E}, {\bs B}, {\bs A}$ are the positive frequency solutions for the electromagnetic fields with the energy $\omega \geqslant 0$, the momentum $k_z$ along the $z$ axis, and the quantized angular number $m \in \Z$, corresponding to the eigenvalue of angular momentum about the axis $z$. In Eq.~\eq{eq:decomposition:G} the radial functions ${\bs G}(\rho)$ are determined by the Maxwell equations~\eq{eq:Maxwell} and by the boundary conditions that will be specified below.

\subsubsection{Boundary conditions}

The spectrum of solutions of the Maxwell equations~\eq{eq:Maxwell} depends on the type of the boundary conditions at the edge of the cylinder at a fixed radial coordinate $\rho = R$. We will consider three kinds of the boundary conditions, corresponding to the boundary made of a perfect electric conductor (an ideal metal), its ``dual'' analogue, a perfect magnetic conductor and finally duality invariant "natural" boundary conditions in infinite space.

We will study an ensemble of rotating photons in a fixed laboratory frame. That means we should have energy and angular
momentum as conserved charges to which we can couple corresponding Lagrange multipliers, the temperature $T$ and
the angular velocity $\Omega$ to define a grand canonical ensemble. 

For the Maxwell field the energy and momentum conservation take the form
\beqn
\frac{\partial \epsilon}{\partial t} + {\bs \nabla} \cdot {\bs P}  = 0\,,
\qquad
\frac{\partial P_l}{\partial t} + \nabla_m \sigma^m\,_l  = 0\,,
\label{eq:conservation:EP}
\eeqn
where the energy, momentum density (Poynting vector) and stress tensor are, respectively, as follows:
\begin{align}
\epsilon &= \frac 1 2 \left( {\bs E}^2 + {\bs B}^2\right)\,\label{energy}\\
{\bs P} &= {\bs E}\times{\bs B}\,,\label{eq:poynting}\\
\sigma_{ml} &= - E_m E_l - B_m B_l + \frac 1 2 g_{ml} \left({\bs E}^2 + {\bs B}^2\right)\label{eq:stress}\,.
\end{align}
Here ${\bs A} \cdot {\bs B} \equiv \sum_{l=1}^3 A_l B_l $ is the scalar product and $l,m = 1,2,3$ are the spatial indices.

In the cylindrical geometry the globally conserved quantities are the energy $\epsilon$, the momentum along the cylinder axis $P_z$ and the $z$ component of the angular momentum $L_z \equiv ({\bs \rho} \times {\bs P})_z = \rho P_\varphi$. As we require from the boundary of the cylinder to respect the conservation of these quantities, Eqs.~\eq{eq:conservation:EP} imply that these quantities are conserved provided both the radial component of the Pointing vector~\eq{eq:poynting} and the radial components of the photon stress tensor~\eq{eq:stress} vanish at $\rho=R$:
\beqn
\begin{array}{rl}
P_\rho(R) &= \phantom{-} E_\varphi(R) B_z(R) - E_z(R) B_\varphi(R) =0\,,
\label{cons_energy} \\[1mm]
\sigma_{\rho\varphi}(R) &= -E_\rho(R) E_\varphi(R) - B_\rho(R) B_\varphi(R) =0
\label{cons_Lphi}\,,\\[1mm]
\sigma_{\rho  z}(R) & {= -E_\rho(R) E_z(R) - B_\rho(R) B_z(R) =0}
\label{cons_Pz}\,.
\end{array} \quad
\label{eq:cons}
\eeqn

Therefore one may distinguish three types of the boundary conditions:
\begin{enumerate}

\item[\bl] Ideal electric conductor. An external electromagnetic field generates dissipationless electric currents in an ideal electric conductor that lead to vanishing normal (with respect to the surface element $\partial S$ of the conductor) component of the external magnetic field $B_\perp$ and two tangential components ${\bs E}_\|$ of the electric field {at the surface boundary}:
\beqn
B_\perp {\biggl|}_{x \in S} = 0\,, \qquad {\bs E}_\| {\biggl|}_{x \in S} = 0\,.
\label{eq:BC:perfect:electric}
\eeqn
In cylindrical coordinates the boundary conditions imposed by the perfect electric conductor~\eq{eq:BC:perfect:electric} have the following form:
\beqn
B_\rho(R) = E_z(R) = E_\varphi(R) = 0\,.
\label{eq:electric:BC}
\eeqn
These conditions ensure conservation of the energy as well as $z$ components of momentum and angular momentum~\eq{eq:cons}.

\item[\bl] Ideal magnetic conductor is a dual analogue of the ideal electric conductor: instead of electric currents, the perfect magnetic conductor hosts dissipationless magnetic currents.\footnote{The perfect-magnetic boundary conditions can be viewed as the electromagnetic analog of the boundary conditions for gluonic field in the MIT bag model for hadrons in QCD.} The magnetic boundary conditions are therefore dual to the electric ones~\eq{eq:BC:perfect:electric}:
\beqn
E_\perp  {\biggl|}_{x \in S} = 0\,, \quad {\bs B}_\|  {\biggl|}_{x \in S} = 0\,.
\label{eq:BC:perfect:magnetic}
\eeqn

The magnetic conductor~\eq{eq:BC:perfect:magnetic} imposes the following conditions on the electromagnetic fields which ensure the physical constraints~\eq{eq:cons}:
\beqn
E_\rho(R) = B_z(R) = B_\varphi(R) = 0\,.
\label{eq:magnetic:BC}
\eeqn
One can readily observe that the perfect electric conductor or perfect magnetic conductor impose the conditions, Eq.~\eq{eq:electric:BC} and \eq{eq:magnetic:BC}, that are mutually  ``dual'' to each other. These boundary conditions will impose either Dirichlet or Neumann boundary conditions {on a scalar radial function of the photon field} depending on its polarization. The electromagnetic duality {transformation}
\beqn
{\bs E} \to - {\bs B}\,, 
\qquad
{\bs B} \to {\bs E}\,,
\label{eq:duality}
\eeqn
exchanges the boundary conditions between the two possible polarizations.

\item[\bl] Unbounded flat space. This is the limit $R\rightarrow \infty$. We impose ``natural'' boundary conditions {by requiring that} the fields and their products should be integrable with the measure $\int_0^\infty \rho d\rho$. These fields can be represented by a Fourier--Bessel integral. In principle, the basis of eigenfunctions for both previously considered boundary conditions can be {also used for the unbounded flat space. However it turns out that it is} slightly more convenient to introduce
in this case an explicitly helicity--preserving basis in terms of left- and right-circularly polarized photon wave functions.

\end{enumerate}

\subsection{Solutions}
\label{sec:solutions}

\subsubsection{Quantization and normalization of electromagnetic fields}

It is convenient to characterize the photon solutions in the interior of the cylinder by transverse electric and transverse magnetic polarization modes. The transverse electric (TE) mode possesses the electric field which is always perpendicular to the axis of the cylinder, $E^\TE_z = 0$. In the transverse magnetic (TM) mode it is the magnetic field which is perpendicular to cylinder's axis, $B^\TE_z = 0$. 

For the quantization of the gauge field it is convenient to choose the Coulomb gauge, where the temporal component of the gauge field is zero and the spatial part of the gauge field has a zero divergence:
\beqn
A_0(x) = 0, \qquad \nab \cdot {\bs A}(x) = 0.
\label{eq:Coulomb:gauge}
\eeqn
Then the photon operator is given by 
\beqn
 {\bs {\hat{A}}}_\mu(x) {=} \sum_{J,\lambda} \frac{{\bs \epsilon}^{(\lambda)}_J} {\sqrt{2 \omega_J}} \left( \cA^{(\lambda)}_{J}(x) {\hat a}^{(\lambda)}_J 
+ \cA^{(\lambda),*}_{J}(x) {\hat a}^{(\lambda)\dagger}_J \right), \quad\
\label{eq:hat:A}
\eeqn
where $\lambda = \TE,\TM$ is the polarization of the photon field and $J$ is a collective notation for other quantum numbers which will be defined below. 

In Eq.~\eq{eq:hat:A} the operators ${\hat a}^{(\lambda)}_J$ and ${\hat a}^{(\lambda)\dagger}_J$ annihilate and, respectively, create a photon with the polarization $\lambda$, the quantum number $J$, and the wavefunction $\cA^{(\lambda)}_{J,\mu}$. These operators obey the standard set of bosonic commutation relations:
\beqs
\beqn
& & \left[ {\hat a}^{(\lambda)}_J ,  {\hat a}^{(\lambda')\dagger}_{J'} \right] = \delta_{\lambda\lambda'} \delta_{JJ'}\,, \\
& & \left[ {\hat a}^{(\lambda)}_J ,  {\hat a}^{(\lambda')}_{J'} \right] = \left[ {\hat a}^{(\lambda)\dagger}_J ,  {\hat a}^{(\lambda')\dagger}_{J'} \right] = 0, \qquad
\eeqn
\eeqs
where $\delta_{JJ'}$ is an identity in the phase space of quantum numbers $J$ with a natural property:
\beqn
\sum_J \delta_{JJ'} = 1 \quad \mbox{for any $J'$}.
\label{eq:identity:phase}
\eeqn

The photonic vacuum state is annihilated by the operators ${\hat a}^{(\lambda)}_J$ for all $\lambda$ and $J$:
\beqn
{\hat a}^{(\lambda)}_J  | 0 \rangle = 0\,.
\eeqn

The photon wavefunctions with a definite polarization~$\lambda$ 
\beqn
{\bs \cA}^{(\lambda)}_J = {\bs \epsilon}^{(\lambda)}_J \cA^{(\lambda)}_J,
\label{eq:A:J}
\eeqn
are defined by the orthonormal vectors 
\beqn
{\bs \epsilon}^{(\lambda)}_J \cdot {\bs \epsilon}^{(\lambda')}_J = \delta_{\lambda\lambda'}\,, \qquad \lambda = \TE, \, \TM,
\eeqn
for each fixed quantum number $J$. For a fixed polarization $\lambda$, the expansion coefficients of the photon operator~\eq{eq:hat:A} are orthonormalized according to the condition:
\beqn
\int d^3 x\, \cA^{(\lambda)*}_{J}({\bs x}) \cA^{(\lambda)}_{J'} ({\bs x}) = \delta_{JJ'}\,.
\label{eq:norm:A}
\eeqn
{In our conventions} there are no sums over repeating indices [e.g., over {the cumulative index} $J$ in Eq.~\eq{eq:A:J}] unless explicitly indicated.

\subsubsection{Explicit solutions at finite radius}

The (positive frequency) expansion coefficients of the photon operator~\eq{eq:hat:A} may be represented as follows,
\beqn
{\bs \cA}^{(\lambda)}(\rho, \varphi, z, t) = e^{- i \omega t + i k_z z + i m \varphi} {\bs \cA}^{(\lambda)}(\rho)\,,
\label{eq:cA:decomposition}
\eeqn
where 
\beqn
\omega = \sqrt{k_z^2 + k_\perp^2}\,,
\label{eq:omega:kz}
\eeqn
is the frequency of the mode, $k_z$ is the momentum along the axis of the cylinder and $m \in \Z$ is the quantum angular momentum associated with the angular rotations about the $z$ axis. The quantization of the transverse (radial) momentum $k_\perp > 0$ in Eq.~\eq{eq:omega:kz} depends on the boundary conditions at the edge of the cylinder.

In the cylindrical coordinates, ${\bs A} \equiv (A_\rho, A_\varphi, A_z)^T$, the radial part of the wave function~\eq{eq:cA:decomposition} is given, for the TE and TM polarizations, respectively, as follows:
\beqs
\beqn
{\bs \cA}^\TE(\rho) & = & \vect{\frac{m f_{\TE}(\rho)}{i\rho}}{\frac{\partial f_\TE(\rho)}{\partial \rho}}{0}\!,
\label{eq:curly:A:TE}
\\[3mm]
{\bs \cA}^\TM(\rho) & = & \vect{ \frac{k_z}{i \omega} \frac{\partial f_\TM(\rho)}{\partial \rho}}{ \frac{m k_z}{\omega} \frac{f_{\TM}(\rho)}{\rho} }{- \frac{k_\perp^2}{\omega} f_\TM(\rho) }\!,
\quad
\label{eq:curly:A:TM}
\eeqn
\label{eq:curly:A}
\eeqs
where the scalar radial functions $f_{\lambda} = f_{\lambda}(\rho)$ obey the following differential equation ($\lambda = \TE, \, \TM$):
\beqn
\frac{1}{\rho} \frac{\partial }{\partial \rho}  \left( \rho \frac{\partial f_\lambda}{\partial \rho} \right) - \frac{m^2}{\rho^2} f_\lambda + k_\perp^2 f_\lambda = 0.
\label{eq:f:diff}
\eeqn
The normalized solutions of Eq.~\eq{eq:f:diff} are proportional to the Bessel functions of the first kind:
\beqn
f_\lambda = C_\lambda J_m(k_\perp \rho)\,, \qquad \lambda = \TE,\,\TM\,,
\label{eq:f:lambda}
\eeqn
where $C_\lambda$ are the normalization constants to be defined below.

In the Coulomb gauge the operators of electric and magnetic fields are given by the series similar to Eq.~\eq{eq:hat:A} where the expansion coefficients can be determined with the help of Eq.~\eq{eq:EB:A}. The electric field modes are proportional to the corresponding gauge field modes~\eq{eq:curly:A}:
\beqn
{\bs \cE}^{(\lambda)} & = & - 
{\partial_t} {\bs \cA}^{(\lambda)} 
= i \omega {\bs \cA}^{(\lambda)},
\label{eq:E:omega}
\eeqn
while the magnetic-field modes ${\bs \cB}^{(\lambda)} = \nab \times {\bs \cA}^{(\lambda)}$ for the $\lambda = \TE,\, \TM$ polarizations are as follows:
\beqs
\beqn
{\bs \cB}^\TE(\rho) & = & \vect{- i k_z \frac{\partial f_\TE(\rho)}{\partial \rho}}{m k_z \frac{f_\TE(\rho)}{\rho}}{[-1mm]{}_{- k_\perp^2 f_\TE(\rho)}},
\label{eq:curly:B:TE}
\\[1mm]
{\bs \cB}^\TM(\rho) & = & \vect{ -i m \omega \frac{f_\TM(\rho)}{\rho}}{\omega \frac{\partial f_\TM(\rho)}{\partial \rho}}{0}.
\label{eq:curly:B:TM}
\eeqn
\label{eq:curly:B}
\eeqs

The cylinder made of an ideal electric conductor or a magnetic conductor imposes, respectively, the boundary conditions~\eq{eq:electric:BC} or \eq{eq:magnetic:BC} on electromagnetic fields of the modes. These constraints can be rewritten as conditions on the radial functions of the corresponding electromagnetic modes:
\beqs
\beqn
\frac{\partial f_\TE(\rho)}{\partial \rho} {\biggl|}_{\rho=R} = f_\TM(R) & = & 0\,, \quad
\bc = E\,,
\qquad \\
\frac{\partial f_\TM(\rho)}{\partial \rho} {\biggl|}_{\rho=R} = f_\TE(R) & = & 0 \,, 
\quad
\bc = M\,. \qquad 
\eeqn
\label{eq:BC:TME}
\eeqs
For shortness, we call the boundary conditions corresponding to the perfect metal~\eq{eq:electric:BC} and the perfect magnetic conductor~\eq{eq:magnetic:BC} as ``electric'' ($\bc = E$) and ``magnetic'' ($\bc = M$) conditions, respectively. A duality of the electric and magnetic boundary conditions with respect to the TE and TM modes is clearly seen in Eq.~\eq{eq:BC:TME}.

The explicit form of the solutions~\eq{eq:f:lambda} indicates that the boundary conditions~\eq{eq:BC:TME} impose the following quantization of the radial momentum $k_\perp$ for the photon polarizations $\lambda$:
\beqn
J'_m (k_\perp R) & = & 0, \quad 
{\vectt{\lambda}{\bc} = \vectt{\TE}{E} , \vectt{\TM}{M}}, \qquad \\
J_m (k_\perp R) & = & 0, \quad 
{\vectt{\lambda}{\bc} = \vectt{\TE}{M} , \vectt{\TM}{E}}, \qquad
\eeqn
where the prime indicates a derivative of the Bessel function with respect to its argument. Thus the walls of the cylinder made of a perfect electric ($\bc = E$) or magnetic ($\bc = M$) conductor quantize the transverse momentum $k_\perp$ of the TE and TM photon modes differently:
\beqs
\beqn
k^{\TE}_\perp = \frac{\kappa'_{ml}}{R}\,, \qquad k^{\TM}_\perp = \frac{\kappa_{ml}}{R}\,, \qquad \bc = E\,,
\label{eq:k:E}\\
k^{\TE}_\perp = \frac{\kappa_{ml}}{R}\,, \qquad k^{\TM}_\perp = \frac{\kappa'_{ml}}{R}\,, \qquad \bc = M\,,
\label{eq:k:M}
\eeqn
\label{eq:k:EM}
\eeqs
where $\kappa_{ml}$ and $\kappa'_{ml}$ (with $m \in \Z$) correspond to the $l$th positive root (with $l =1,2, \dots \in \N$) of the Bessel function $J_m(x)$ and its derivative $J'_m(x)$, respectively: 
\beqn
J_m(\kappa_{ml}) = 0\,, \qquad J'_m(\kappa'_{ml}) = 0\,.
\label{eq:J:prime:0}
\eeqn
According to Eq.~\eq{eq:omega:kz} the corresponding frequencies $\omega$ of the electromagnetic modes are:
\beqs
\beqn
\omega_J = \sqrt{k_z^2 + \frac{(\kappa'_{ml})^2}{R^2}}, \quad 
{\vectt{\lambda}{\bc} = \vectt{\TE}{E} , \vectt{\TM}{M}}, \quad \\
\omega_J = \sqrt{k_z^2 + \frac{(\kappa_{ml})^2}{R^2}}\,,  \quad 
{\vectt{\lambda}{\bc} = \vectt{\TE}{M} , \vectt{\TM}{E}}. \quad
\eeqn
\label{eq:omega:J:EM}
\eeqs

In a cylinder the photonic modes of a definite polarization $\lambda$ are labeled by the collective quantum number~\eq{eq:J}.
\beqn
J = \{k_z, m, l\}\,, \quad k_z \in \R, \quad m \in \Z, \quad l \in \N.
\label{eq:J}
\eeqn
An integration over all three momenta ${\bs k}$ in a phase space of plane waves in an unrestricted space is reduced, in the cylinder, to the sum over the collective quantum number~$J$:
\beqn
\int \frac{d^3 k}{(2\pi)^3} \quad \longleftrightarrow \quad  \sum_J = \frac{1}{\pi R^2} \int \frac{d k_z}{2 \pi} \sum_{m \in \Z} \sum_{l = 1}^\infty.
\label{eq:sum:J}
\eeqn
This sums appears, for example, in Eq.~\eq{eq:hat:A}.

According to Eq.~\eq{eq:identity:phase} the identity in the phase space of the modes with a given polarization $\lambda$ is as follows:
\beqn
\delta_{JJ'} = 2 \pi^2 R^2 \delta(k_z - k_z') \delta_{mm'} \delta_{ll'}.
\eeqn

An explicit calculation of the orthonormalization condition~\eq{eq:norm:A},
\beqn
\int_0^R d \rho \rho \, f_\TE^2(\rho) = \int_0^R d \rho \rho \, f_\TM^2(\rho) = \frac{R^2}{2 k_\perp^2},
\eeqn
gives us the coefficients~$C^\bc_\lambda$:
\beqs
\beqn
C_\TE^{E} = C_\TM^{M} & = & \frac{R}{\sqrt{ (\kappa'_{ml})^2 - m^2} \, |J_{m}(\kappa'_{ml})|},
\label{eq:C:TE:E}\\
C_\TE^{M} = C_\TM^{E} & = & \frac{R}{\kappa_{ml} \left| J_{m+1}(\kappa_{ml}) \right|},
\label{eq:C:TE:M}
\eeqn
\label{eq:C:TE}
\eeqs
in the radial functions~\eq{eq:f:lambda} of the photon polarization modes $\lambda=\TE,\,\TM$ obeying the $\bc=E,M$ boundary conditions. Here we used the integral orthogonality relations of the Bessel functions~\eq{eq:Bessel:int:1} and \eq{eq:Bessel:int:2}, as well as the recurrence relations~\eq{eq:Bessel:recurrence}. Notice that $\kappa'_{ml} > |m|$.

The TM and TE modes are always orthogonal to each other,
\beqn
\int\limits_0^R d \rho \rho \, {\bs \cA}^\TE_J(\rho) {\cdot} {\bs \cA}^\TM_J(\rho) \propto \left[ m f_\TE(\rho) f_\TM(\rho) \right] {\biggl|}_0^R \equiv 0,\quad 
\eeqn
due to the boundary conditions~\eq{eq:BC:TME} and the fact that $m J_m(0) \equiv 0$ for all $m\in \Z$.

The conserved charges of interest in this basis are the total energy and the angular momentum {which are eigenvalues of the Hamiltonian and the angular momentum operators. In our normalization the normal ordered expressions of these operators are, respectively, as follows}:
\begin{align}
\cH &= \int d^3x \frac 1 2 :( {\bs E}^2 + {\bs B}^2):\, = \sum_{J,\lambda} \omega_{J}^{(\lambda)}\, \hat a^{(\lambda)\dagger}_J \hat a^{(\lambda)}_J \label{eq:hamiltonian}\,,\\
L_{\varphi} &= \int d^3x :P_\varphi:\, = \sum_{J,\lambda} m \,\hat a^{(\lambda)\dagger}_J \hat a^{(\lambda)}_J \label{eq:angularmomentum}\,.
\end{align}

\subsubsection{Modes in an unbounded space with $R \to \infty$}

As a final point in this section we will discuss {the} limit of an unbounded space $R\rightarrow \infty$. First let us note that without imposing any boundary conditions we have
\begin{equation}
{\bs \nabla} \times \bcA_J^{(TE,TM)} = \omega \bcA_J^{(TM,TE)}\,.
\end{equation}
We can therefore introduce eigenvectors of the curl operator 
\begin{equation}
\bcA^{\pm}_J = \bcA_J^{TE}  \pm \bcA_J^{TM} 
\end{equation}
with the eigenvalues
\begin{equation}
{\bs \nabla}\times \bcA^{\pm}_J = \pm \omega \bcA^{\pm}_J\,.
\end{equation}
In terms of electric and magnetic fields these modes fulfill {the relations}
\begin{equation}
{\bs B}_J^{\pm} = \mp i {\bs E}_J^{\pm}\,,
\end{equation}
{which show that these modes correspond to} left- and right-circularly polarized electromagnetic fields. The gauge potential can now be quantized in this basis as follows:
\begin{equation}\label{eq:helicityA}
{\bs \cA} = \sum_J \frac{\sqrt{2}}{\sqrt{\omega_J}} \left( {\bcA}_J^{(+)} \alpha_J^{(+)} + \bcA_J^{(-)*} \alpha_J^{(-)\dagger}\right)\,.
\end{equation}
Similarly to the finite-radius cases, the radial scalar functions $f_\lambda$ are still proportional to a Bessel function~\eq{eq:f:lambda}. However the radial momentum $k_\perp$ is not quantized in the absence of the boundaries. The wave functions are still normalized according to the condition:
\beqn
\int d^3 x\, {\bs \cA}^{(\lambda)*}_{mk}(x) {\bs \cA}^{(\lambda')}_{m',k'}(x) = \delta_{\lambda\lambda'} \delta_{JJ'}
\eeqn
with the collective quantum number $J = (m, k, k_\perp)$ and
\beqn
\delta_{JJ'} & = & 4 \pi^2 \delta_{mm'} \delta(k_z - k'_z) \frac{\delta(k_\perp - k'_\perp)}{k_\perp}, \\
\sum_J & = & \int_{-\infty}^{+\infty} \frac{d k_z}{2\pi} \sum_{m\in\Z} \int_0^\infty \frac{k_\perp d k_\perp}{2\pi}.
\label{eq:sum:J:infty}
\eeqn
The normalization constant in this unbounded case is $C=\frac{1}{\sqrt{2}k_\perp}$. Since the wave functions for both circular polarizations obey the same boundary conditions (see below), the normalization constant is the same for both polarizations. The complex field ${\bs \cE} = {\bs E} + i {\bs B}$ is then just ${\bs\cE} = {\bs \nabla} \times {\bs \cA}$.

Quantization is achieved by 
\begin{equation}
[ \alpha_{J}^{(\lambda)\dagger}, \alpha_{K}^{(\mu)} ] = \delta_{JK}\delta^{\lambda\mu}\,.
\end{equation}
The wave functions form an orthonormal system 
\begin{equation}
\int d^3x\, \bcA_K^{(\lambda)*} \cdot \bcA_L^{(\mu)}   = \delta_{K,L} \delta^{\lambda,\mu}\,.
\end{equation}
We note that ${\bs E} \pm i {\bs B}$ are eigenvectors of the duality transformation $({\bs E},{\bs B}) \rightarrow ({\bs B}, - {\bs E})$ with eigenvalues $\pm i$.
The Hamiltonian is ${\cH}= \frac 1 2 {\bs \cE} \cdot {\bs \cE}^\dagger$. Both polarization modes have the same frequencies. Therefore
the Hamiltonian is
\begin{equation}
\cH = \sum_J \omega_J \left(\alpha^{(+)\dagger}_J\alpha^{(+)}_J + \alpha^{(-)\dagger}_J\alpha^{(-)}_J \right)\,,
\label{eq:Hamiltonian}
\end{equation}
with $\omega_J^2 = k_z^2 + k_\perp^2$ {as in Eq.~\eq{eq:omega:kz}}. 
{The projection of the angular momentum on the $z$ axis} can be computed from the expression of the
Poynting vector ${\bs P} = \frac i 2 {\bs \cE}\times {\bs \cE}^\dagger$ as
\begin{equation}
L_\varphi = \sum_J m \left(\alpha^{(+)\dagger}_J\alpha^{(+)}_J + \alpha^{(-)\dagger}_J\alpha^{(-)}_J \right)\,,
\label{eq:angular:momentum}
\end{equation}

\subsubsection{Helicity and zilch}

Back in the 60's Lipkin found a new conserved charge for free Maxwell theory which he
called the "zilch" \cite{Lipkin}. Soon afterwards Kibble pointed out that there are infinitely many such zilch currents \cite{Kibble}. 

The basic observation is the following: if $({\bs E},{\bs B})$ and $({\bs H},{\bs G})$ are doublets of fields obeying free Maxwell's equations
\begin{eqnarray}
\nabla \cdot {\bs E} =  \nabla \cdot {\bs B} =  \nabla \cdot {\bs H} =  \nabla \cdot {\bs G} =0 \,,\\
\nabla \times {\bs B} - \frac{\partial {\bs E}}{\partial t} = \nabla \times {\bs  E} + \frac{\partial {\bs B}}{\partial t} =0 \,,\\
\nabla \times {\bs H} - \frac{\partial {\bs G}}{\partial t} = \nabla \times {\bs  G} + \frac{\partial {\bs H}}{\partial t} =0\,,
\end{eqnarray}
then the following expression 
\begin{eqnarray}\label{eq:defzilch}
\zeta = {\bs H} \cdot {\bs B} + {\bs G} \cdot {\bs E}\\ \label{eq:defzilchcur}
{\bs J}_\zeta = -{\bs H} \times {\bs E} + {\bs G} \times {\bs B}\,,
\end{eqnarray}
fulfill a conservation law
\begin{equation}
\frac{\partial \zeta}{\partial t} + \nabla \cdot {\bs J}_\zeta =0 \,.
\label{eq:zilch:conservation}
\end{equation}

If we identify ${\bs H} \, {\to}\, {\bs A}$ with the vector-potential and ${\bs G}\, {\to}\,{\bs C}$  and the dual vector potential in the Coulomb gauge $\nabla \cdot {\bs A} = \nabla \cdot {\bs C} =0$, then
\begin{eqnarray}
{\bs B} =  \nabla \times {\bs A} =  \frac{\partial { \bs C}}{\partial t },
\\
{\bs E} = - \frac{\partial { \bs A}}{\partial t}
= \nabla \times {\bs C}.
\end{eqnarray}
The conserved charge (\ref{eq:defzilch}) in this case is the optical helicity.
The inconvenience with these definitions is that they do depend on the gauge choice. The vector and dual vector potential
define a zilch current only in the Coulomb gauge~{\eq{eq:Coulomb:gauge}}! 

The original zilch current of Lipkin is distinguished in that it is gauge invariant and local. It can be defined by
taking
\begin{eqnarray}
{\bs H} = \nabla \times {\bs B} \,,\\
{\bs G} = \nabla \times {\bs E}\,.
\end{eqnarray}

If one allows for non-local expressions one can define the $k$-zilch currents by taking
\begin{eqnarray}
{
{\bs H}_s = \Delta^{-s}{\bs \nabla} \times {\bs B}\,, 
}\\
{
{\bs G}_s = \Delta^{-s} {\bs \nabla} \times {\bs E}\,,
}
\end{eqnarray}
where $\Delta$ is the Laplace operator. If one uses the Coulomb gauge~{\eq{eq:Coulomb:gauge}} then the $1$-zilch {($s=1$)} becomes local in terms of the vector
potentials $\bs A$, $\bs C$ 
and it coincides with the ``optical helicity'' {given in a relativistic form in Eq.~\eq{eq:optical:helicity}}. The $1$-zilch is also distinguished in {a sense} that it is the only one of the $s$-zilches that has a correct  dimension of a conserved current, i.e. dimension 3. In contrast the gauge invariant local zilch current of Lipkin, the $0$-zilch, has dimension 5, {and is often associated with ``the optical chirality flow''~\cite{ref:Cameron,Elbistan:2016xzq}.}

For a finite radius case, the perfect electric~\eq{eq:electric:BC} and magnetic~\eq{eq:magnetic:BC} conductor boundary conditions do not respect the zilch. The helicity (or zilch) influx 
\beqn \label{eq:helicityinflux}
{\bs J}_{h} \cdot {\bs n} & \equiv & J_{h,r}(R)   \\
& = & \left(E_\varphi A_z - E_z A_\varphi + C_\varphi B_z - C_z B_\varphi \right) |_{\rho=R},
\nonumber
\eeqn
does not vanish identically at the boundary.

In the helicity eigenstate basis in the unbounded domain the helicity and the zilch can be expressed by the complex fields
\begin{align}
h &= \frac 1 4({\bs \cA}^\dagger \cdot {\bs \cE} + {\bs \cA} \cdot  {\bs \cE}^\dagger) \label{eq:helicitydensity}\,,\\
\zeta &= \frac 1 4( {\bs \cG}^\dagger \cdot {\bs \cE} + {\bs \cE} \cdot {\bs \cG}^\dagger)\label{eq:zilchdensity}\,.
\end{align}
The normal ordered integrated total charges (helicities and zilches) are, respectively, as follows:
\begin{align}
Q_h &= \int d^3x\, :h:= \sum_J \left( \alpha^{(+)\dagger }_J\alpha^{(+)}_J - \alpha^{(-)\dagger}_J \alpha^{(-)}_J \right), \label{eq:totalhelicity}\\
Q_\zeta &=  \int d^3x\, :\zeta: = \sum_J \omega_J^2 \left( \alpha^{(+)\dagger}_J \alpha^{(+)}_J - \alpha^{(-)\dagger}_J \alpha^{(-)}_J\right) \label{eq:totalzilch}\,.
\end{align}
As expected, helicity in the Coulomb gauge counts the number of right-circularly polarized photons minus the number
of left-circularly polarized photons. The gauge invariant zilch charge weights these numbers with the squares of
the frequencies and is therefore a good gauge invariant observable and local measure of helicity \cite{tangcohen}.

The expression of the helicity and zilch currents in terms of the complex fields are
\begin{align}
{\bs J}_h &= \frac{i}{4} \left( {\bs\cA}\times {\bs\cE}^\dagger - {\bs\cA}^\dagger\times {\bs\cE}\right)\,,\\
{\bs J}_\zeta &= \frac{i}{4} \left( {\bs\cE} \times {\bs\cG}^\dagger - {\bs\cE}^\dagger \times{\bs \cG}\right)\,.
\end{align}

\subsection{Rotations}

We will study the rotating ensemble in {a vacuum defined with respect to} a fixed laboratory frame. In this case rotation is implemented by defining the statistical operator
\begin{equation}\label{eq:densitymatrix}
\rho = \frac{1}{Z} e^{-\beta(\cH - \Omega L_\varphi)}\,,
\end{equation}
where $\beta = 1/T$ is the inverse temperature, $\Omega$ is the angular {frequency corresponding to a uniform rotation with angular velocity ${\bs \Omega} = \Omega \, {\bs {\mathrm{e}}_z}$about the $z$ axis}, {$\cH$ is the Hamiltonian~\eq{eq:Hamiltonian} and $L_\varphi$ is the projection of the angular momentum operator on the rotation axis~\eq{eq:angular:momentum}}. Without loosing generality we assume that the cylinder rotates counterclockwise with $\Omega \geqslant 0$. 

Rotating ensembles of relativistic field theories are notoriously ill-defined whenever the tangential velocity at radius $\rho$ exceeds the speed of light. A well defined ensemble is therefore possible only as long as $R\Omega<1$, where the speed of light $c=1$ in our conventions. This makes it immediately clear that the unbounded domain a constant angular velocity does not give rise to a consistent statistical ensemble. As noted however long ago by Vilenkin, in the case of fermions it is possible to extract meaningful results for the statistical average of the current at the center of rotation. As we will discuss in detail, for photons even this {property is not realized} beyond a lowest order in $\Omega$.

In principle one can study the ensemble both in a co-rotating and in a laboratory (non-rotating) frame and define two different vacua. A nonrotating vacuum has been considered by Vilenkin~\cite{Vilenkin:1980zv} while the rotating vacuum has been studied by Iyer~\cite{Iyer:1982ah}. One may show that both approaches are equivalent provided the system is bounded in such a way that the velocity of the rigidly rotating body does not exceed the speed of light so that the causality is respected. Technically, the nonrotating (Vilenkin) vacuum is equivalent to the rotating (Iyer) vacuum if the energy of each eigenmode in the laboratory frame $\varepsilon$ and in the co-rotating frame $\tilde\varepsilon$ satisfy the relation $\varepsilon\tilde\varepsilon > 0$. This relation always holds in the case if the causality is respected. The causality is lost in a rigidly rotating unbounded space, in which $\varepsilon\tilde\varepsilon < 0$ for certain modes and, consequently, the nonrotating and rotating vacua are not equivalent~\cite{Ambrus:2014uqa}. The unbounded rotating systems may have several pathologies related to instabilities and rotation-induced Unruh effect~\cite{ref:pathologies:1,ref:pathologies:2}. Further discussions, in particular on a difference between fermionic and bosonic states, may be found in Ref.~\cite{Ambrus:2014uqa}.

The thermal expectation value of an operator $\cO$ {for a uniformly rotating ensemble is}:
\beqn
\langle \hat\cO(x)\rangle_{T,\Omega} = \sum_{J,\lambda} n_B(T,\Omega ;J,\lambda) \langle\hat\cO(x)\rangle_J\,,
\label{eq:O:vev:lab}
\eeqn
where $\langle\hat\cO(x)\rangle_J \equiv \avr{J|\cO|J}$ corresponds to the value of the operator $\cO$ for a photon in the state {characterized by the polarization $\lambda$ and the kinetic quantum numbers $J$~\eq{eq:J}}, and
\beqn
n_B(T,\Omega;J,\lambda) = \frac{1}{e^{(\omega_J^{(\lambda)}- m \Omega)/T} - 1}
\label{eq:n:T}
\eeqn
is the Bose-Einstein distribution function at nonzero temperature $T$ and angular velocity $\Omega$. 

In order to calculate the expectation values of the normal ordered operators of interest we collect the
mode expansions of the different fields
\begin{align}
{\bs A} &= \sum_{J,\lambda} \frac{1}{\sqrt{2\omega_J^{(\lambda)}}}
\left({\bs \cA}_J^{(\lambda)} \hat{a}_J^{(\lambda)} + {\bs \cA}_J^{(\lambda),*} \hat{a}_J^{(\lambda)\dagger}\right)\,,\\
{\bs C} &= \sum_{J,\lambda} \frac{i}{\sqrt{2\omega_J^{(\lambda)}}}
\left({\bs {\widetilde{\cA}}}_J^{(\lambda)} \hat{a}_J^{(\lambda)} - {\bs {\widetilde{\cA}}}_J^{(\lambda),*} \hat{a}_J^{(\lambda)\dagger}\right)\,,\\
{\bs E} &= \sum_{J,\lambda} \frac{i\sqrt{\omega_J^{(\lambda)}}}{\sqrt{2}}
\left({\bs \cA}_J^{(\lambda)} \hat{a}_J^{(\lambda)} - {\bs \cA}_J^{(\lambda),*} \hat{a}_J^{(\lambda)\dagger}\right)\,,\\
{\bs B} &= \sum_{J,\lambda} \frac{\sqrt{\omega_J^{(\lambda)}}}{\sqrt{2}}
\left({\bs {\widetilde{\cA}}}_J^{(\lambda)} \hat{a}_J^{(\lambda)} + {\bs {\widetilde{\cA}}}_J^{(\lambda),*} \hat{a}_J^{(\lambda)\dagger}\right)\,,
\end{align}
\begin{align}
{\bs H} &= \sum_{J,\lambda} \frac{(\omega_J^{(\lambda)})^{3/2}}{\sqrt{2}}
\left({\bs \cA}_J^{(\lambda)} \hat{a}_J^{(\lambda)} - {\bs \cA}_J^{(\lambda),*} \hat{a}_J^{(\lambda)\dagger}\right)\,,\\
{\bs G} &= \sum_{J,\lambda} \frac{i(\omega_J^{(\lambda)})^{3/2}}{\sqrt{2}}
\left({\bs {\widetilde{\cA}}}_J^{(\lambda)} \hat{a}_J^{(\lambda)} + {\bs {\widetilde{\cA}}}_J^{(\lambda),*} \hat{a}_J^{(\lambda)\dagger}\right)\,,
\end{align}
where the dual wave functions ${\bs {\widetilde{\cA}}}$ are defined via the relation $\nabla \times {\bs \cA}^{(\lambda)}_J = \omega {\bs {\widetilde{\cA}}}^{(\lambda)}_J$.

Now we can compute the thermal averages of the following normal ordered operators:
\begin{align}
\mathrm{optical\,helicity:}\ \, & J^0_h = \frac 1 2 :({\bs A}\cdot {\bs B} + {\bs C} \cdot {\bs E}):\,,\\
&{\bs J}_h = \frac 1 2 :( {\bs E} \times {\bs A} + {\bs C} \times {\bs B} ):\,,\\
\mathrm{zilch:}\ \, & J_\zeta^0  = \frac 1 2 :({\bs H}\cdot {\bs B} + {\bs G} \cdot {\bs E}): \,,\\
& {\bs J}_{\zeta} = \frac 1 2 :({\bs E} \times {\bs H}  + {\bs G}\times{\bs B}   ):\,,\\
\mathrm{Poynting\ vector:}\ \, &{\bs J}_{\epsilon}  = :{\bs E} \times {\bs B}:\,.
\end{align}
We note that all these expression are duality invariant: $({\bs E},{\bs B}) \rightarrow (-{\bs B}, {\bs E})$, $({\bs C}, {\bs A}) \rightarrow (-{\bs A},{\bs C})$ etc. They are taken normal ordered (with creation operators placed to the left) and we only need the one-particle expectation values to evaluate the thermal averages. 

 {
It is worth mentioning that we quantize the gauge field $A_\mu$ in the Coulomb gauge~\eq{eq:Coulomb:gauge} formulated in the laboratory frame. This gauge condition is not satisfied by the fields in the corotating frame $A'_\mu$ which are related to the ones in the laboratory frame by the linear transformation ${\bs A}' = {\bs A}$ and $A_0' = A_0 - \Omega \rho A_\varphi$. The spatial part of the Coulomb gauge is thus respected by the corotating gauge fields, ${\bs \nabla}' \cdot {\bs A}' = 0$, while the temporal component of the gauge field in the corotating frame is nonzero, $A'_0 \neq 0$, for both TE and TM photon polarizations~\eq{eq:curly:A}. However, since all observables of interest are formulated in the laboratory frame, and the vacua for both laboratory and rotating frame are the same, the quantization should be done in the Coulomb gauge~\eq{eq:Coulomb:gauge} with respect to the gauge fields in the laboratory frame. Moreover, the uniform rotation affects the expectation values of the observables in laboratory frame via the Bose-Einstein distribution function~\eq{eq:n:T},  which depends on the photon energy spectrum in the corotating frame, $\omega_J' =\omega_J - m \Omega$. Since the latter is a gauge-independent quantity, the choice of the gauge in the corotating frame has no effect on the expectation values of the observables.
}

The optical helicity, zilch and their currents have the single particle expectation values
\begin{align}
\bra{J} J^0_{h,\zeta} \ket{J} & =  2 k m (\omega_J)^{1 - 2 s}\frac{f_J f_J'}{\rho}\,,\\
\bra{J} {\bs J}_{h,\zeta} \ket{J} &= \vect{0}{ k k_\perp^2 (\omega_J)^{-2 s} f_J f'_J}{m(\omega_J)^{2 - 2 s}\left(1+\frac{k^2}{\omega_J^2}\right) \frac{f_J f'_J}{\rho}} 
\label{eq:J:modes}
\end{align}
where $s=1$ for the optical helicity and $s=0$ for the zilch. We note that these quantities fulfill the Ward identity
\begin{equation}
\omega \bra{J} J^0_{h,\zeta} \ket{J}  - \frac{m}{\rho} \bra{J}  J^\varphi_{h,\zeta} \ket{J} - k\bra{J}  J^z_{h,\zeta} \ket{J}  = 0\,,
\end{equation}
{associated with the zilch conservation~\eq{eq:zilch:conservation} and with a similar conservation relation for the helicity.}

The expectation value of the Poynting vector is
\begin{align}
\bra{J} \vec{J}_\epsilon \ket{J} & = \vect{0}{\frac{m}{\rho} k_\perp^2 f_J^2}{k ( \frac{m^2}{\rho^2}f_J^2 + f_J'^2)} .
\end{align}
For simplicity of notation we have suppressed the polarization index in the above. The expression are formally
the same for both polarizations. 

Since the energy $\omega_J$ is an even function of the momentum in $z$-direction $k$ all thermal expectation values
of expression linear in $k_z$ vanish upon integration. {The linearity in $k_z$} immediately tells us that $h=\zeta=J^z_{\epsilon}= J_{h}^\varphi=J_{\zeta}^{{\varphi}}=0$ {in addition to the obvious absence of the radial currents $J^\rho_{\epsilon}= J_{h}^\rho=J_{\zeta}^\rho=0$}.

{Furthermore}, from the identity
\begin{equation}
2 m\, \int_0^R \rho d\rho\, \frac{f f'}{\rho} = m f(R)^2\,,
\label{eq:integral}
\end{equation}
it follows that {only those modes that obey the Neumann boundary conditions on the radial photon functions $f_J$} give us a nonzero net helicity and zilch currents!
{The correspondence between the boundary conditions on the radial photon functions, the photon polarizations and the type of the boundary conditions can be found in Eq.~\eq{eq:BC:TME}.}

{In general case}, the thermal expectation values can not be evaluated analytically {and therefore we proceed to their} numerical evaluation. For the numerical
summation it is convenient to write the energy and angular momentum {densities} as follows:
\begin{align}
R^4 \epsilon & = \frac{1}{\pi^2} \sum_{m,l,\lambda} \int_0^\infty dq\, \frac{\nu^{(\lambda)}_J }{e^{\frac{\nu^{(\lambda)}_J-m R\Omega}{RT} }-1} \,
\label{eq:energy:num}\\
R^3 L_\varphi &= \frac{1}{\pi^2}\sum_{m,l,\lambda} \int_0^\infty dq\, \frac{m}{e^{\frac{\nu^{(\lambda)}_J-m R\Omega}{RT} }-1}\,.
\label{eq:angular:num}
\end{align}
{In order to adapt these quantities for a numerical evaluation we used a shorthand notation $\nu^{(\lambda)}_J \equiv R\omega^{(\lambda)}_J$ for the dimensionless energy, characterized by the cumulative index $J$ of Eq.~\eq{eq:J} and by the polarization $\lambda = \TE/\TM$ according to the type of the boundary condition~\eq{eq:omega:J:EM}. } 

\begin{figure}[!thb]
\begin{center}
\includegraphics[scale=0.55,clip=true]{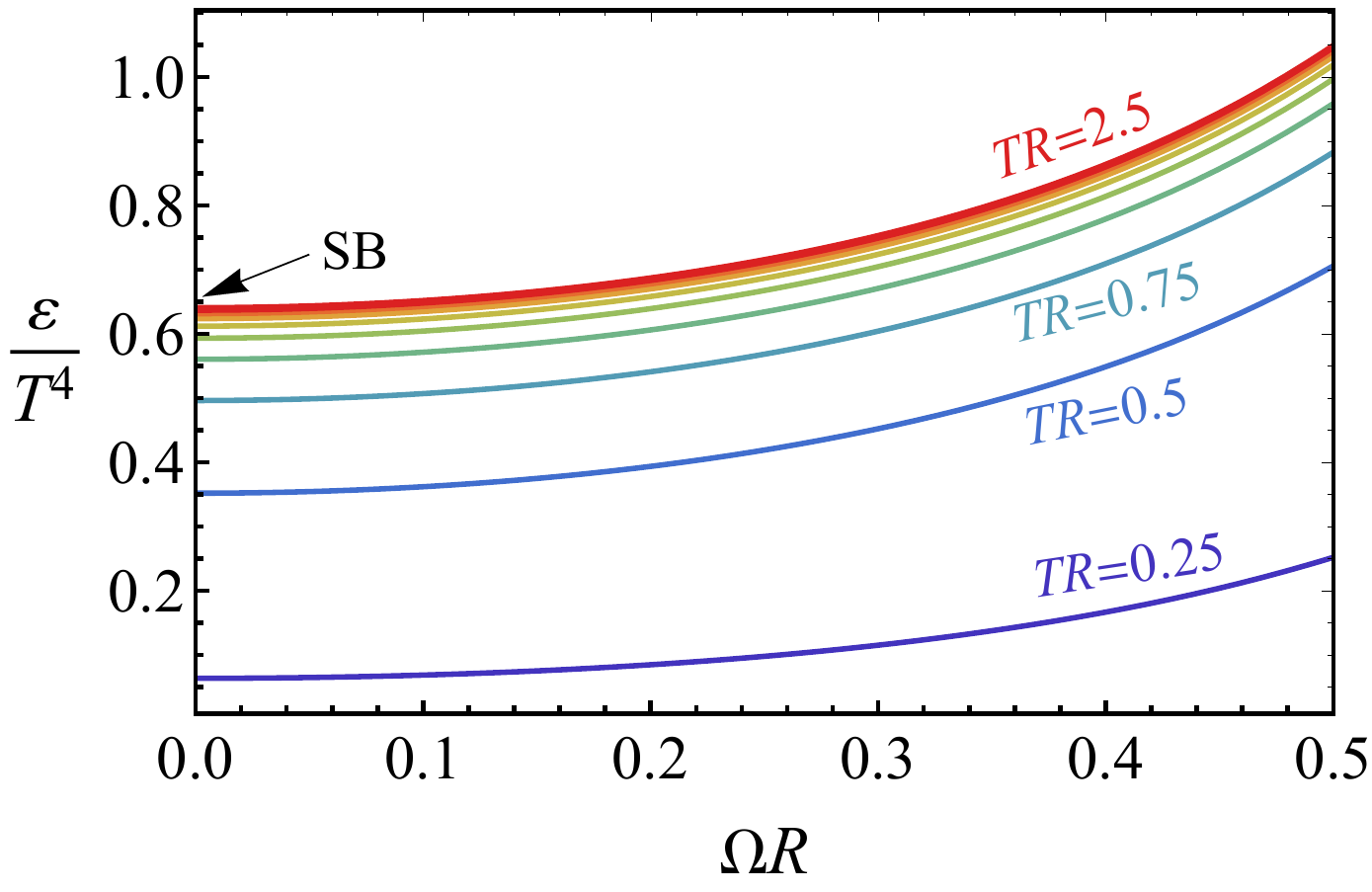}\\[5mm]
\includegraphics[scale=0.55,clip=true]{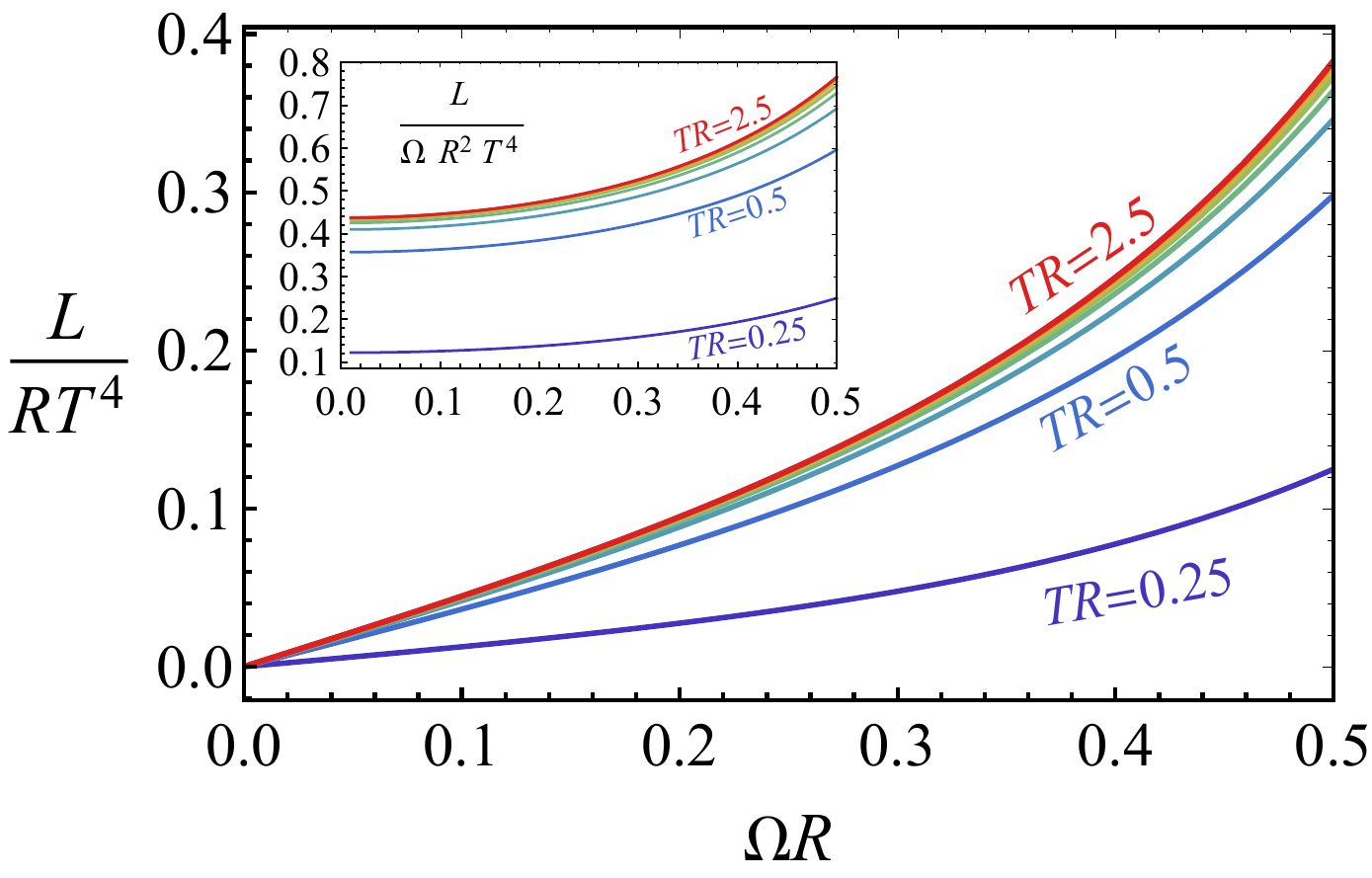}
\end{center}
\vskip -4mm 
\caption{The {energy~\eq{eq:energy:num} and angular momentum~\eq{eq:angular:num}} densities as function of the dimensionless tangential velocity at the boundary $R\Omega$. We restrict ourselves to relatively low angular velocities $R\Omega<0.5$ to facilitate numerical evaluation. The plots show that behavior of the dimensionless quantities $\epsilon/T^4$ (top) and $L/(RT^3)$ (bottom) for a temperature range from $TR=0.25$ up to $TR=2.5$ in steps of $\delta (T R) = 0.25$.  As it can be seen from the plots, both these dimensionless quantities collapse on a universal high temperature curve. The arrow in the upper plot marks the Stefan-Boltzmann value $\epsilon = \frac{T^4\pi^2}{15}$. At temperature $T R=2.5$ the system is already within 3\% of this value at zero rotation. {The inset in the lower figure shows the moment of inertia~\eq{eq:I} with the corresponding Stefan-Boltzmann value.}} 
\label{fig:EL}
\end{figure}

In Fig.~\ref{fig:EL} we show appropriately normalized energy~\eq{eq:energy:num}, angular momentum~\eq{eq:angular:num} and moment of inertia
\beqn
I(\Omega,T) = \frac{L_\varphi(\Omega,T)}{\Omega},
\label{eq:I}
\eeqn
as functions of the angular frequency $\Omega$ for various fixed temperatures $T$.

The components of the total helicity and zilch currents {along the axis of rotation, given by integration over local currents~\eq{eq:J:modes} over the crosssection of the  cylinder,} 
\beqn
{
J^{z,\mathrm{tot}}_s = \int_0^R \rho d \rho \int_0^{2\pi} d \varphi \, J_s(\rho,\varphi)\,,
}
\label{eq:J:tot}
\eeqn
are as follows:
\begin{equation}
J^{z,\mathrm{tot}}_s R^{2-2s} {=} \frac 1 \pi \sum_{m,l} \int\limits_0^\infty d q  \frac{1+\frac{q^2}{(\nu^{(\lambda)}_J)^2}}{(\kappa'_{ml})^2{-}m^2}  \frac{m (\nu^{(\lambda)}_J)^{2-2s}}{e^{\frac{\nu^{(\lambda)}_J - m R\Omega}{RT} }-1},\
\label{eq:J:tot:s}
\end{equation}
where 
$s=1$ corresponds to the helicity and $s=0$  to the zilch currents (also denoted, respectively, as $s = h$ and  $s = \zeta$ below). In Eq.~\eq{eq:J:tot:s} we choose the polarization $\lambda$ taking into account the fact~\eq{eq:integral} that only the modes with the Neumann boundary conditions on the radial photon functions $f_J(\rho)$ may contribute.

\begin{figure}[!thb]
\begin{center}
\includegraphics[scale=0.55,clip=true]{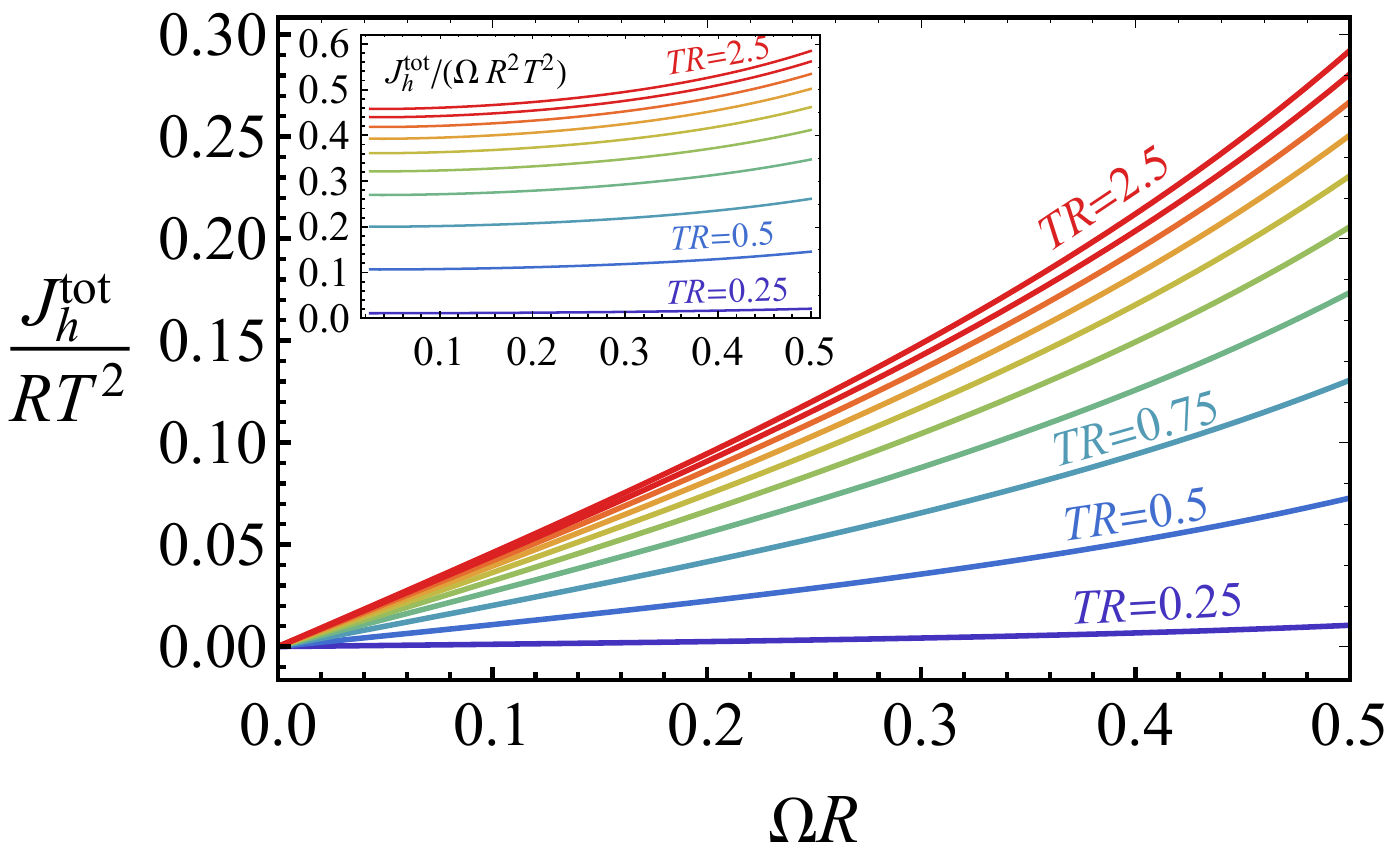}\\[3mm]
\includegraphics[scale=0.55,clip=true]{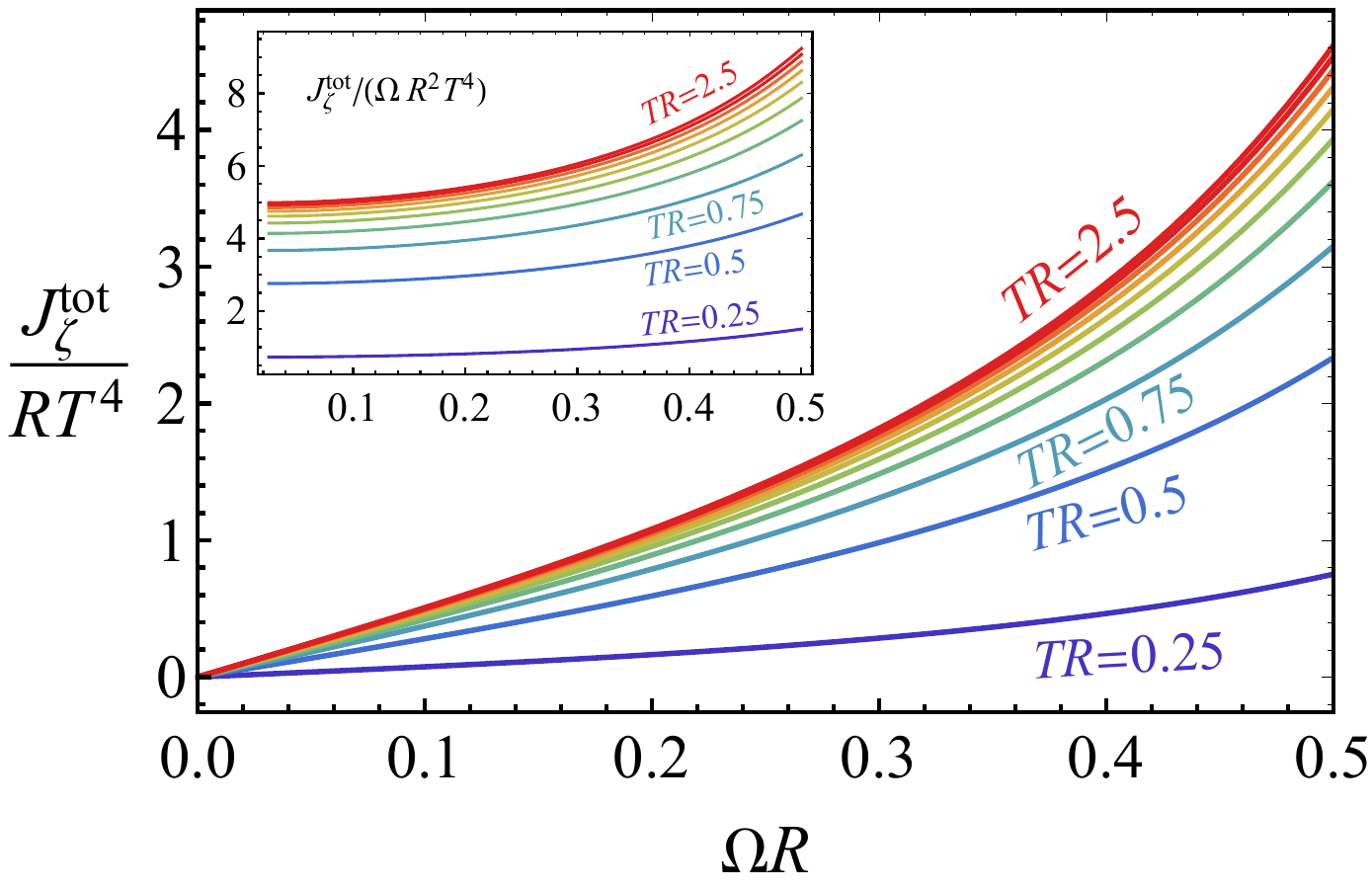}
\end{center}
\vskip -4mm 
\caption{The {total helicity ($s=1$) and zilch ($s=0$) currents~\eq{eq:J:tot:s}} as function of the dimensionless tangential velocity at the boundary $R\Omega$. The plots show that behavior of the dimensionless quantities $J^{z,\mathrm{tot}}_h/(RT^2)$ and $J^{z,\mathrm{tot}}_\zeta/(RT^4)$ for  {the same  temperature range as in Fig.~\ref{fig:EL}. The insets show the currents divided by the frequency $\Omega$.}}  
\label{fig:TotHZ}
\end{figure}

In Fig.~\ref{fig:TotHZ} we show the total helicity and zilch currents~\eq{eq:J:tot:s} which are increasing function of both temperature and angular momentum. Its important to remember that the total currents, {contrary to the infinite-volume expression}, are non-vanishing only because of the Neumann boundary condition {imposed on radial photon functions}. The net flux of helicity and zilch is therefore to be interpreted as an effect of the duality breaking boundary conditions. Qualitatively both quantities exhibit increasing flux with increasing angular velocity.

\begin{figure}[!thb]
\begin{center}
\includegraphics[scale=0.525,clip=true]{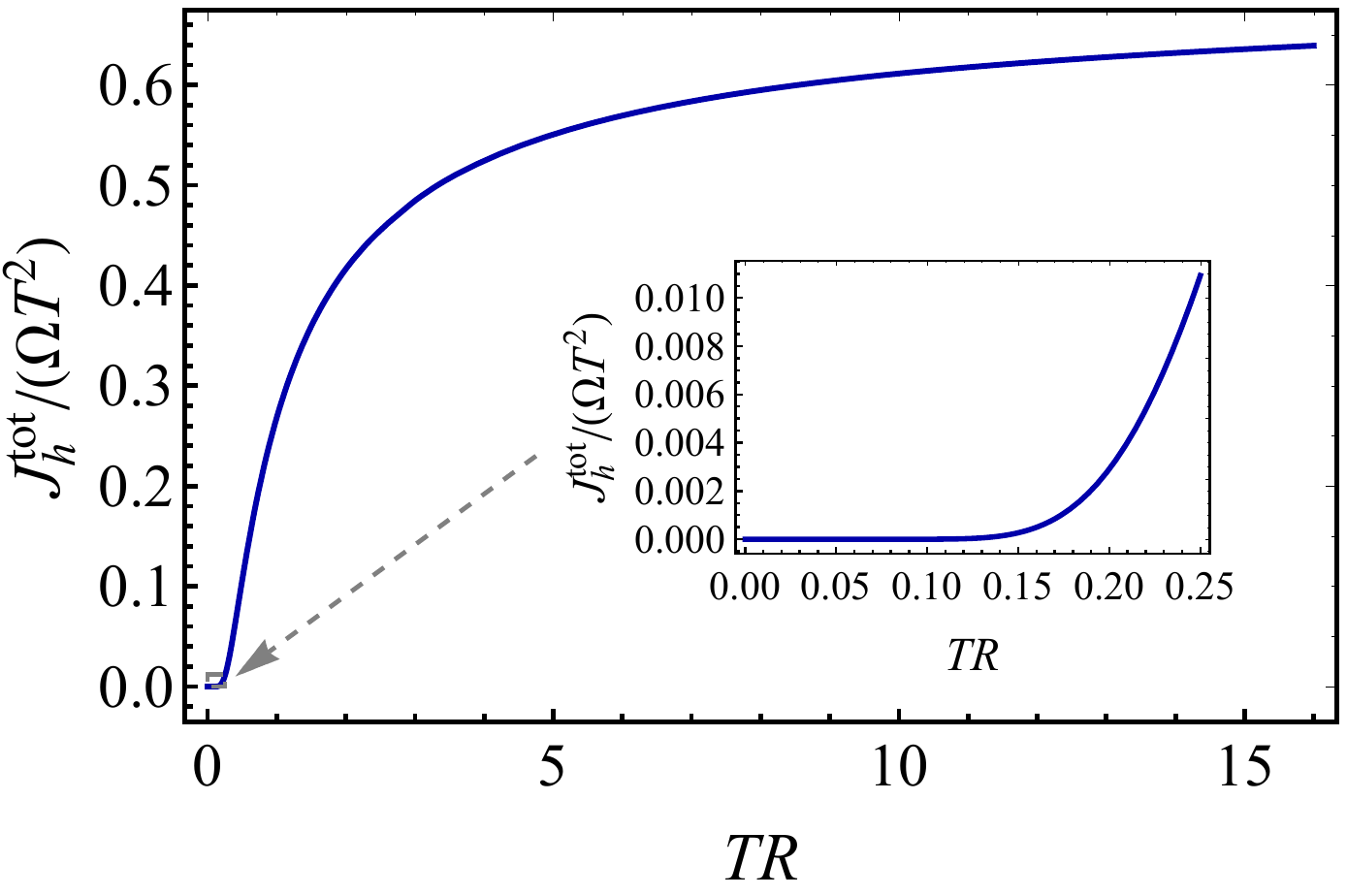}\\[3mm]
\includegraphics[scale=0.525,clip=true]{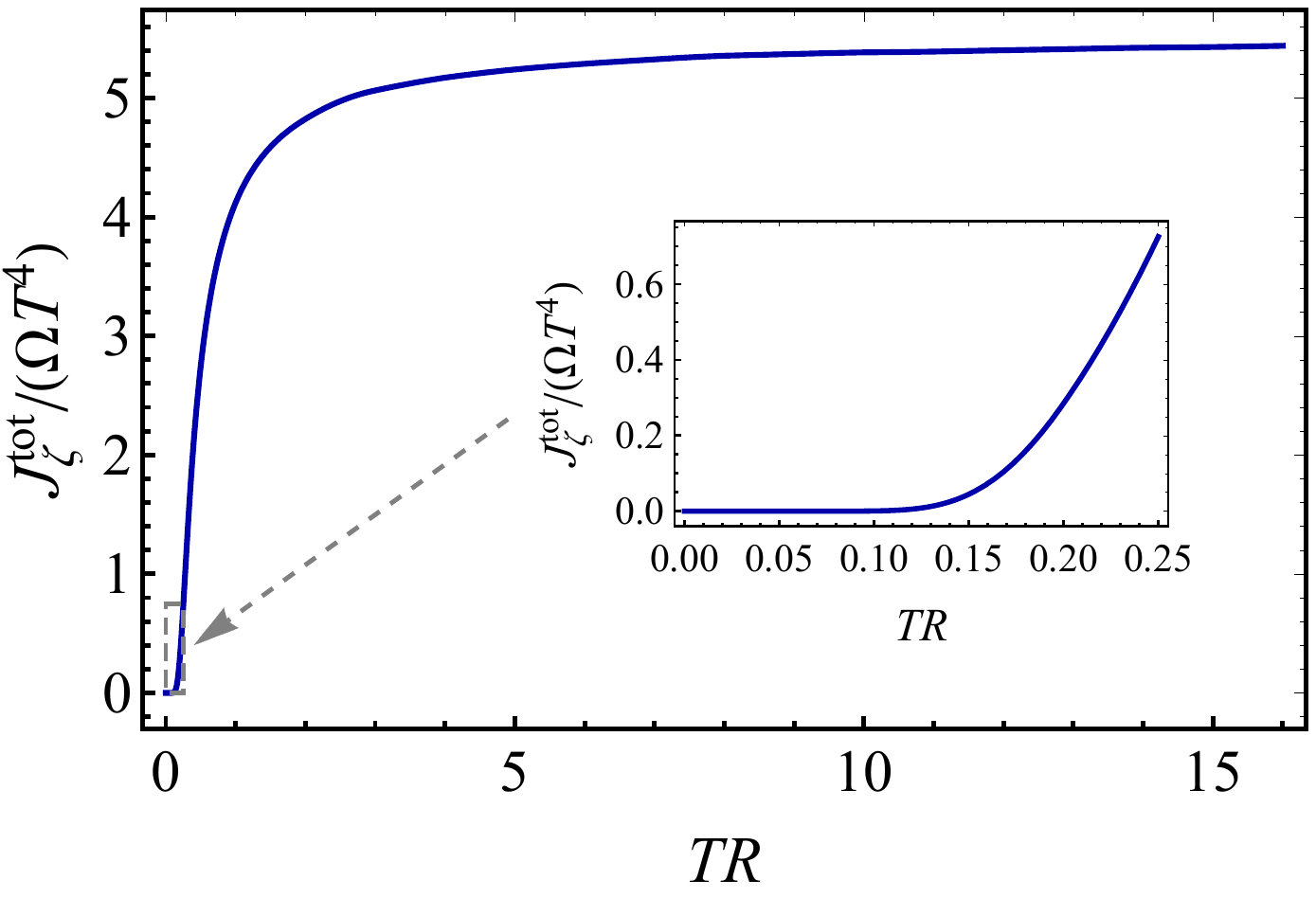}
\end{center}
\vskip -4mm 
\caption{ {The dimensionless strength of the helicity ($s=1$) and zilch ($s=0$) currents~\eq{eq:C:s} in the limit of small angular frequencies $\Omega \to 0$ as function of temperature~$T$. The insets show the exponential onset of both currents at small temperatures.}} 
\label{fig:strengths}
\end{figure}

 {It is well seen in Fig.~\ref{fig:TotHZ} that in the limit of small angular frequencies, $\Omega \to 0$, both the total helicity and the total zilch currents~\eq{eq:J:tot:s} exhibit a linear dependence on $\Omega$:
\beqn
J^{z,\mathrm{tot}}_s = C_s(T) T^{4-2s} \Omega  + O\left(\Omega^2\right)\,.
\label{eq:C:s}
\eeqn
In Fig.~\ref{fig:strengths} we show the dimensionless coefficients $C_s$ as the function of temperature $T$. Both quantities vanish exponentially in the limit of small temperatures $T \to 0$, while in the limit of high temperature they approach the values 
\beqn
C_h(T \to \infty) \approx 0.65, \qquad C_\zeta(T \to \infty) \approx 5.42, 
\label{eq:Cs:num}
\eeqn
respectively.
}

Finally using the series of the Bessel functions, $J_m(x) = (x/2)^m + O(x^{m+2})$, it follows that the helicity and zilch current densities at the axes of rotation $\rho=0$ receive only contributions from the angular momenta $m=\pm 1$:
\begin{align}
J^z_{s}(0) R^{2-2s} &= \frac{1} {\pi^2}
\sum_{l,\lambda} 
 \int\limits_0^\infty dq 
(\nu_J^{(\lambda)})^{(2-2s)} \left( 1 + \frac{q^2}{(\nu_J^{(\lambda)})^2}\right) \nonumber\\
& \hskip -15mm \cdot C_\lambda^2 \left( \frac{\kappa_{1,l}^{\lambda}}{2R} \right)^2 
\left[\frac{1}{e^{\frac{\nu_J^{(\lambda)}- R\Omega}{RT} }-1}  -\frac{1}{e^{\frac{\nu_J^{(\lambda)}+ R\Omega}{RT} }-1}\right],
\label{eq:Jz:s}
\end{align}
where again $s=1$ and $s=0$ correspond to the helicity and the zilch, respectively. {The $m = \pm 1$ eigennumbers $\kappa_{1,l}^{\lambda} \equiv \kappa_{-1,l}^{\lambda}$ for both polarizations $\lambda = \TE/\TM$ can be read off from Eqs.~\eq{eq:k:EM} and \eq{eq:J:prime:0}, and the normalization coefficients $C_\lambda$ are given in Eq.~\eq{eq:C:TE}.}

\begin{figure}[!thb]
\begin{center}
\includegraphics[scale=0.525,clip=true]{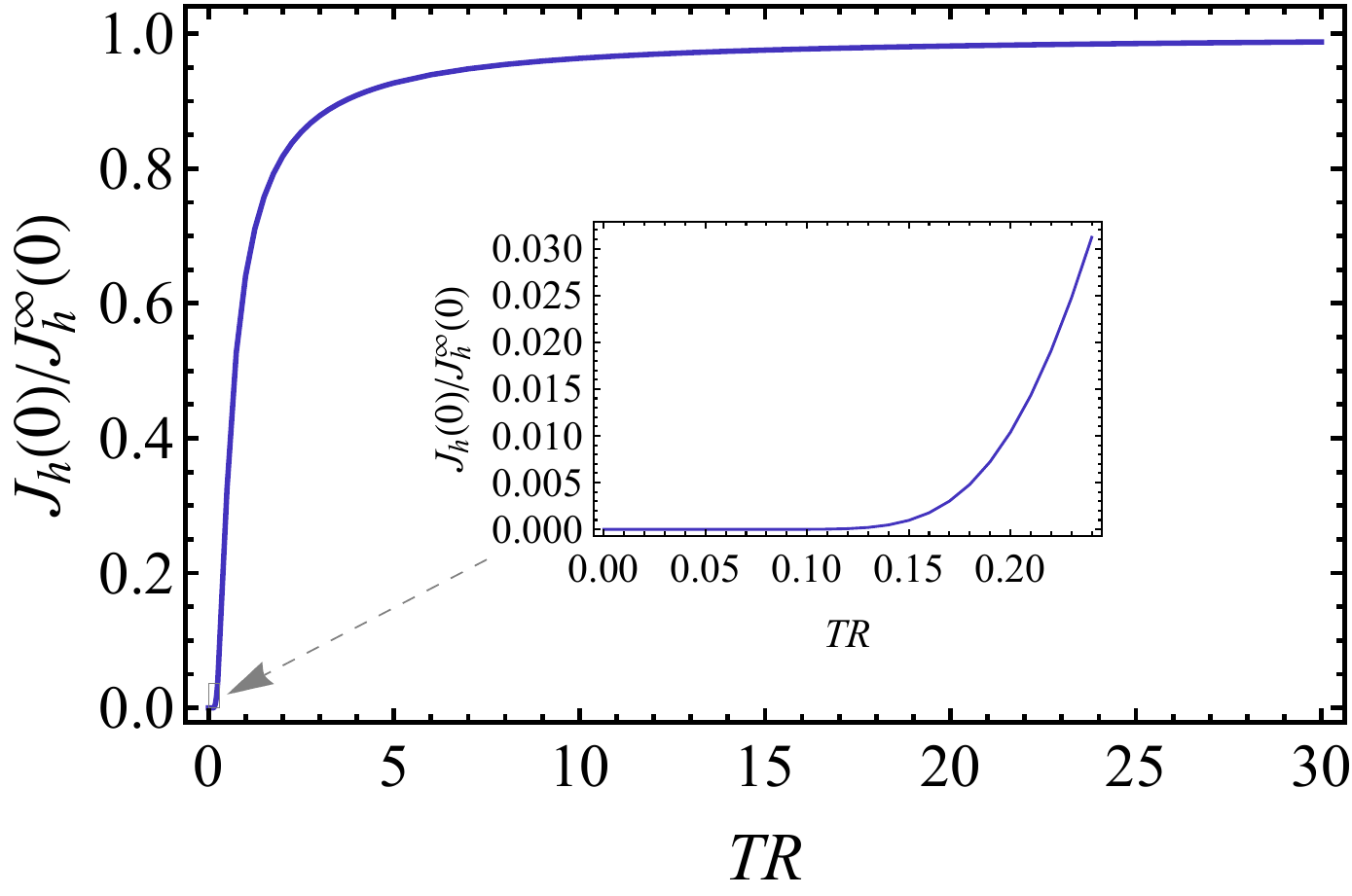}\\[3mm]
\includegraphics[scale=0.525,clip=true]{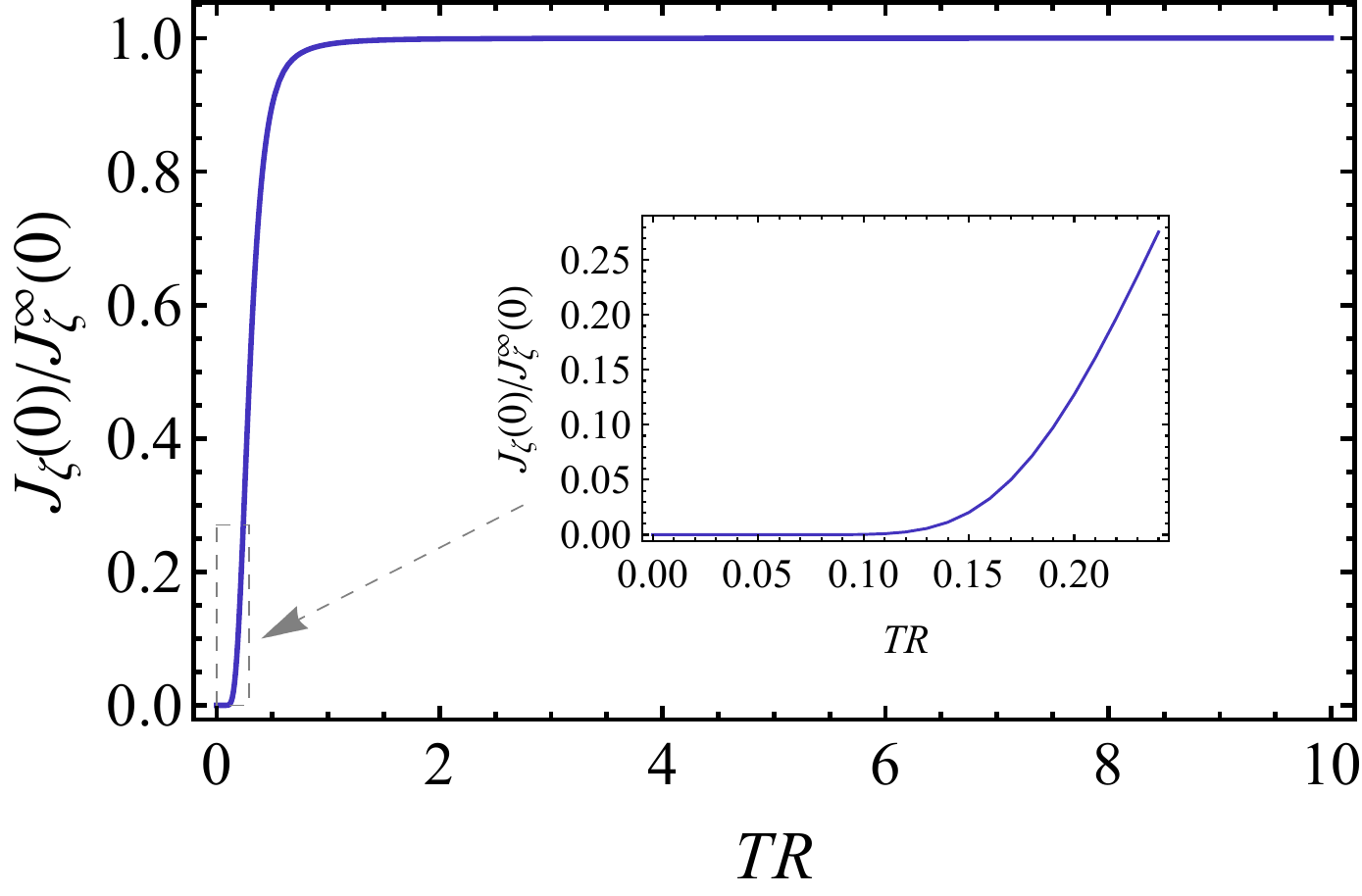}
\end{center}
\vskip -4mm 
\caption{Values of the helicity {($s=1$)} and zilch {($s=0$)} current densities~{\eq{eq:Jz:s}} at the axes of rotation $\rho=0$ for very low angular velocities. The result is plotted as a function of $TR$ and as fraction of the result in the unbounded domain (\ref{eq:helicitycentercur})
and (\ref{eq:zilchcentercur}) respectively. The insets show the currents at small temperatures.} 
\label{fig:CenterCurs}
\end{figure}

The helicity and zilch currents~\eq{eq:Jz:s} at the axis of rotation $\rho = 0$ are shown in Fig.~\ref{fig:CenterCurs} as function of temperature~$T$. We plot these currents in a limit of slow rotations $\Omega \to 0$ and normalized them to the corresponding results obtained in the unbounded domain (\ref{eq:helicitycentercur}) and (\ref{eq:zilchcentercur}), to be discussed in the next section. The high temperature limit approaches the value of the linear truncation in $\Omega$ in the unbounded domain. Its interesting that this convergence is faster for the zilch current than for the helicity current. The insets show the exponential onset of the currents for small temperatures.

\subsection{Unbounded domain}
We will now study the problem of the generation of the helicity and zilch currents at the center of rotation in an unbounded domain. The analogous problem for chiral fermions is known to give a well defined expression that coincides to lowest order in the angular momentum with the predictions from anomaly induced transport theory, the chiral vortical effect. It also predicts terms of higher order in $\Omega$ but their status is somewhat less clear. We will follow the strategy that worked for chiral fermions as close a possible.

The thermal expectation value of 
{
the helicity current
in the unbounded domain} is formally
\begin{align}
\left\langle {{J^z_h}}
{(\rho)}\right\rangle_{T,\Omega}^{{{\infty}}} & =  \int\limits_{\Omega^+}^\infty\frac{dk_\perp}{2\pi}\int\limits_{-\infty}^\infty \frac{dk}{2\pi} \sum_m 
n_B(\omega-m\Omega,T) \nonumber \\
& 2 m \left(1+\frac{k^2}{\omega^2}\right) \frac{ J_m(k_\perp \rho) J'_m(k_\perp \rho)}{\rho} \,.
\label{eq:J:inf:1}
\end{align}
where {$n_B(\varepsilon,T) = [\exp(\varepsilon/T)-1]^{-1}$ is the occupation number and the eigenenergy $\omega$ is given in Eq.~\eq{eq:omega:kz}}. 
Both photon polarizations contribute the same in the current~\eq{eq:J:inf:1}. Since the current~\eq{eq:J:inf:1}
should be understood as the limit $R\rightarrow \infty$ of the finite radius theory there is in principle
a lower limit on the $k_\perp$ integration. At any finite radius we have $\Omega R < 1$ and $k_\perp = \frac{\kappa_{ml}}{R}$ with $\kappa_{ml}>m$. Therefore we always have $k_\perp > \Omega$ in Eq.~\eq{eq:J:inf:1}. 

One observation is that the total current vanishes also in the unbounded domain. {Indeed we integrate the current~\eq{eq:J:inf:1} over the (infinite) crosssection of the cylinder as in Eq.~\eq{eq:J:tot} and then use the identity~\eq{eq:integral} to show that the contribution of every eigenmode $f_J(\rho) = J_m(k_\perp \rho)$ is proportional to $J^2_m(k_\perp R)$ which vanishes in the infinite-volume limit $R \to \infty$.}

If we concentrate on the other hand on
the center of rotation $\rho=0$ we find that only the modes with $m=\pm 1$ contribute.
We can also change the integration variable from $k_\perp$ to $\omega$ and find then
\begin{align}\label{eq:centercurrent}
\left\langle {{J^z_h}}
(0)\right\rangle_{T,\Omega}^{{{\infty}}} & = \frac{1}{8\pi^2} \int_{\Omega^+}^\infty d\omega \int_{-\omega}^\omega {d k}
 \left(\omega+\frac{k^2}{\omega}\right) \nonumber\\
& \left[\frac{1}{e^{(\omega-\Omega)/T}-1} -  \frac{1}{e^{(\omega+\Omega)/T}-1} \right]\,.
\end{align}
We can now expand the integral to lowest order in powers of $\Omega/T$ and find 
\begin{equation}
\label{eq:helicitycentercur}
\left\langle 
{J^z_h}
(0)\right\rangle_{T,\Omega}^{{{\infty}}}  = \frac{4 T^2 \Omega}{3\pi^2} \int_0^\infty dx \frac{x}{e^x-1} = \frac{2 T^2}{9} \Omega\,.
\end{equation}

One can also try to proceed by ignoring the lower bound on the integration {over the frequency $\omega$ in the first integral} in Eq.~(\ref{eq:centercurrent}). This leads
to
\begin{align}
\left\langle 
{J_h^z}
(0)\right\rangle_{T,\Omega}^{{{\infty,(\mathrm{formal})}}} &= \frac{T^3}{3\pi^2} \int_0^\infty dx x^2 \nonumber\\
& \left[ \frac{1}{e^{x-\Omega/T} -1} - \frac{1}{e^{x+\Omega/T} -1}\right]\,.
\end{align}
In order to evaluate this we can use the integral representation of the polylogarithms together with the Jonqui{\'e}re
inversion relation
\begin{align}
Li_n(z)& = \frac{1}{\Gamma(n)} \int_0^\infty \frac{t^{n-1}}{e^t/z-1}\,,\nonumber\\
Li_n(z) +(-1)^n Li_n(1/z) &= -\frac{(2\pi i)^2}{n!} B_n\left(\frac 1 2 \pm \frac{\ln(-z)}{2\pi i}\right)\,.\nonumber
\end{align}
Here $B_n(x)$ is the n-th Bernoulli polynomial and the sign is chosen according to $z\not\in [0,1]$ or $z\not\in]1,\infty]$.
This leads to the following {formal} expression:
\begin{equation}
\left\langle 
{J_h^z}(0)\right\rangle_{T,\Omega}^{{{\infty,(\mathrm{formal})}}} = \frac{2 T^2}{9} \Omega \pm i \frac{T}{3 \pi}\Omega^2 - \frac{1}{9\pi^2}\Omega^3\,,
\label{eq:J0:formal}
\end{equation}
which is clearly unphysical beyond the leading order in {the angular momentum} $\Omega$. The reason is that the integrand in {Eq.~\eq{eq:centercurrent}} has always at least one
pole at frequency $\omega=|\Omega|$. The analogous integrals for fermions are well defined since the Fermi-Dirac distribution
does not present a singularity. However even in the fermionic case the higher order terms do not seem to be universal~\cite{Stone:2018zel}.

The same considerations go through for the zilch current as well. The only difference is an additional insertion of $\omega^2$ under the integral {in Eq.~\eq{eq:J:inf:1}}. We only quote the {infinite-volume result for the on-axis zilch current obtained in the} linear order in $\Omega$:
\begin{equation}
\label{eq:zilchcentercur}
\left\langle 
{J^z_\zeta} (0)\right\rangle_{T,\Omega}^{{{\infty}}} = \frac{8 \pi^2 T^4}{45} \Omega\,.
\end{equation}

\section{Discussion and Conclusion}

We have studied helicity and zilch photonic currents in free Maxwell theory induced by rotation in a bounded cylindrical domain. An important role is played by the conditions {imposed on photons at the boundary of the cylinder}. We have chosen two types of the boundary conditions corresponding to perfect electric and perfect magnetic conductors as both these conditions guarantee that the influx of energy and angular momentum vanishes at the boundary. The values of  {the helicity and zilch photonic currents} for both types of boundaries are the same because these boundary conditions are exchanged under a discrete electric-magnetic duality transformation while all expression of interest are duality invariant.

In looking out for an analogue of the well--known CVE of chiral fermions we studied the current densities at the axis of rotation. A universal value can reasonably be expected to arise only in the high temperature limit in which the boundary conditions play no role for {the physics of photons at the center of rotation}. Indeed we found that in the high--temperature limit the result for small angular velocity converges to the result obtained to linear order in $\Omega$ in the unbounded domain. {However, a direct} calculation in the unbounded domain is plagued with the difficulty that the integrals over the Bose-Einstein distributions are well defined only for {sufficiently small angular velocities}, $\Omega < 1/R$. This fact means that the angular velocity has to go faster to zero than $1/R$ {goes}. Consequently, the {formal result for the helicity current at the axis of rotation}, obtained in the unbounded domain~{\eq{eq:J0:formal}}, does not seem to be physically meaningful as this procedure gives a complex value for an expectation value of a hermitian operator. 

{On the contrary, a truncation of the expression for the current to a lowest order in $\Omega$ in unbounded domain provides us with a still meaningful physical result~{\eq{eq:helicitycentercur}} because it exactly corresponds to a value to which the central current densities converge in the-high temperature limit in the bounded domain. In this sense (the leading order truncation of the high-temperature limit) one can indeed speak of a chiral vortical effect for photons in an unbounded domain.}

{It is worth comparing} our numerical result for the central helicity current (\ref{eq:helicitycentercur}) to the {existing} derivations of the photonic CVE {in the literature}~\cite{Avkhadiev:2017fxj, Yamamoto:2017uul, Zyuzin}. We note that Refs.~\cite{Avkhadiev:2017fxj,Zyuzin} consider the magnetic helicity current 
{(with the results, in our notations, $J_h^z = T^2 \Omega/6$ and $J_h^z = (\epsilon \mu - 1) T^2 \Omega/12$, respectively)}
and only Ref.~\cite{Yamamoto:2017uul} studies a semiclassical evaluation of the optical helicity current {(which gives $J_h^z = T^2 \Omega/3$)}. {Notice that all these expressions for the helicity currents differ from each other (in particular, the helicity current of Ref.~\cite{Zyuzin} is zero in the vacuum $\epsilon = \mu = 1$).} In any case, our value {for the helicity current~\eq{eq:helicitycentercur}} differs from the results obtained in all these works. 

{The disagreement in the literature} is probably not surprising since the helicity current is not a gauge invariant object and, therefore, it can not be considered a good physical observable. On the other hand Lipkin's zilch current is a local and gauge invariant {quantity}. {The zilch current at the axis of the rotating cylinder in the high-temperature limit is given in Eq.~\eq{eq:zilchcentercur}}. It would be interesting to evaluate the expectation value of the central zilch current via Kubo formulas or in a semiclassical treatment to compare to our result~(\ref{eq:zilchcentercur}).

 {
Summarizing, we have found the Zilch Vortical Effect (ZVE) which generates the helicity and zilch currents along the axis of rotation of a hot gas of photons. We have calculated these currents in a wide domain of temperatures and angular frequencies (Figs.~\ref{fig:TotHZ} and \ref{fig:CenterCurs}) in a causality-preserving setup. For a photon gas in a fixed-size cavity, the currents vanish exponentially in the limit of low temperature. At high temperature and low angular frequency of rotation, the currents at the axis of rotation are given by Eqs.~\eq{eq:helicitycentercur} and \eq{eq:zilchcentercur} while the total currents are estimated in Eqs.~\eq{eq:C:s} and \eq{eq:Cs:num}.
}

 {
Both the helicity and zilch currents show qualitatively similar behavior. They constitute a part of an infinite tower of conserved charges, zilches, of free electromagnetic field. Thus, in a general sense, the ZVE is responsible for an infinite tower of anomalous transport effects in a rotating photon gas. 
}

\acknowledgements
K.L. would like to thank the participants of the workshop "Open problems and Opportunities in chiral fluids" Santa Fe, 17-19 July 2018, especially M.~Stone and A.~Sadofyev for interesting and useful discussions.  M.Ch. is grateful to D.~E.~Kharzeev for useful discussions. We also thank M.~Elbistan for correspondence and comments on the manuscript. This work has been supported by the PIC2016FR6/PICS07480, FPA2015-65480-P(MINECO/FEDER) and Severo Ochoa Excellence Program grant SEV-2016-0597. The work of M.Ch. was partially supported within the state assignment of the Ministry of Science and Education of Russia (Grant No. 3.6261.2017/8.9). A.C. acknowledges financial support through the MINECO/AEI/FEDER, UE Grant No. FIS2015-73454-JIN.
\appendix
\section{Cylindrical coordinates}
\label{appendix:cylindrical}

In the cylindrical coordinates a vector
\beqn
{\bs a} = a_\rho \e_\rho + a_\varphi \e_\varphi + a_z \e_z\,,
\label{eq:vector:a}
\eeqn
is represented via the orthonormalized basis vectors of the cylindrical system:
\beqn
\e_\rho = \vect{\cos \varphi}{\sin\varphi}{0} \!,\
\e_\varphi = \vect{-\sin \varphi}{\cos\varphi}{0} \!,\
\e_z = \vect{0}{0}{1}\!, \qquad
\eeqn
where $\varphi$ is the azimuthal angle in the $(x, y)$ plane, related to the cartesian coordinates as follows:
\beqn
x = \rho \cos\varphi\,, \qquad y = \rho \sin\varphi\,.
\eeqn

The basic operations of the vector calculus are:\\
\noindent
-- The scalar product
\beqn
{\bs a} \cdot {\bs b} = a_\rho b_\rho + a_\varphi b_\varphi + a_z b_z\,.
\eeqn
\noindent
-- The vector product
\beqn
\left( {\bs a} \times {\bs b} \right)_\rho & = & \left( a_\varphi b_z - b_\varphi a_z \right) \,, \\
\left( {\bs a} \times {\bs b} \right)_\varphi & = & \left( a_z b_\rho - b_z a_\rho \right) \,, \\
\left( {\bs a} \times {\bs b} \right)_z & = & \left( a_\rho b_\varphi - b_\rho a_\varphi \right) \,.
\eeqn
\noindent
-- The curl (rotor) operation:
\beqn
\left( \nab \times a \right)_\rho & = & \frac{1}{\rho} \frac{\partial a_z}{\partial \varphi} - \frac{\partial a_\varphi}{\partial z}\,, \\
\left( \nab \times a \right)_\varphi & = & \frac{\partial a_\rho}{\partial z} - \frac{\partial a_z}{\partial \rho}\,, \\
\left( \nab \times a \right)_z & = & \frac{1}{\rho} \frac{\partial (\rho a_\varphi)}{\partial \rho} - \frac{1}{\rho} \frac{\partial a_\rho}{\partial \varphi}\,.
\eeqn
\noindent
-- The divergence: 
\beqn
\nab \cdot {\bs a} & = & \frac{1}{\rho} \frac{\partial (\rho a_\rho)}{\partial \rho} +  \frac{1}{\rho} \frac{\partial a_\varphi}{\partial \varphi} + \frac{\partial a_z}{\partial z}\,.
\eeqn
\noindent
-- The gradient:
\beqn
\nab f & = & \frac{\partial f}{\partial \rho} {\bs \e}_\rho + \frac{1}{\rho} \frac{\partial f}{\partial \varphi} {\bs \e}_\varphi + \frac{\partial f}{\partial z} {\bs \e}_z\,.
\eeqn
\noindent
-- The Laplacian:
\beqn
\Delta f & = & \frac{1}{\rho} \frac{\partial }{\partial \rho} \biggl(\rho \frac{\partial f}{\partial \rho} \biggr)+ \frac{1}{\rho^2} \frac{\partial^2 f}{\partial \varphi^2} + \frac{\partial^2 f}{\partial z^2}\,.
\eeqn

\section{Some properties of Bessel functions}
\label{appendix:Bessel}

The Bessel functions satisfy the following recurrence relations:
\beqs
\beqn
J_{m-1}(x)  + J_{m+1}(x) & = & \frac{2 m}{x} J_{m}(x), \\
J_{m-1}(x)  - J_{m+1}(x) & = & 2 J'_{m}(x).
\eeqn
\label{eq:Bessel:recurrence}
\eeqs

For arbitrary parameters $a$ and $b$ one gets:
\beqn
& &\int_0^1 dx x J_m(a x) J_m(b x) \nonumber \\
& & \hskip 15mm = \frac{b J_m(a) J_{m-1}(b) - a J_m(b) J_{m-1}(a)}{a^2-b^2},\\
& & \int_0^1 dx x^2 J_m(a x) J'_m(a x) = \frac{1}{2a} J_{m-1}(a) J_{m+1}(a). \qquad
\label{eq:Bessel:int:3}
\eeqn

If $a = \kappa_{ml}$ and $b = \kappa_{ml'}$ are zeros of the Bessel function, $J_m(\kappa_{ml}) = J_m(\kappa_{ml'}) = 0$, then 
\beqn
\int_0^1 dx x J_m(\kappa_{ml} x) J_m(\kappa_{ml'} x) = \frac{\delta_{ll'}}{2} J_{m+1}^2(\kappa_{ml})\,.
\label{eq:Bessel:int:1}
\eeqn
If $a = \kappa'_{ml}$ and $b = \kappa'_{ml'}$ are zeros of a derivative of the Bessel function, $J'_m(\kappa'_{ml}) = J'_m(\kappa'_{ml'}) = 0$, then 
\beqn
& & \int_0^1 dx x J_m(\kappa'_{ml} x) J_m(\kappa'_{ml'} x) \nonumber \\
& & \hskip 15mm = \frac{\delta_{ll'}}{2} \left[ J_m^2(\kappa'_{ml}) - J_{m+1}^2(\kappa'_{ml})  \right]\,.
\label{eq:Bessel:int:2}
\eeqn

For real positive $k$ and $k'$ one gets:
\beqn
\int_0^{\infty} d\rho \rho \left[ \frac{m^2}{\rho^2} J_m(k \rho) J_m(k' \rho) + k k' J'_m(k \rho) J'_m(k' \rho) \right] &&\nonumber \\
\equiv k^2  \int_0^{\infty} d\rho \rho J_m(k \rho) J_m(k' \rho) = k \delta(k-k'). \qquad &&
\label{eq:Bessel:int:4}
\eeqn

Finally we note that for large index the asymptotic expansions of the zeros are
\begin{align}
\kappa_{m1} &= m + 1.8558 m^{1/3} + O(m^{-2/3}) \,\\
\kappa_{m1}' &\sim m + 0.8086 m^{1/3} + O(m^{-2/3} \,.
\end{align}
this makes the divergence of the thermodynamic partition function for $\Omega R > 1 $ explicit.


\begin{thebibliography}{99}

\bibitem{Kharzeev:2013ffa}
  D.~E.~Kharzeev,
  ``The Chiral Magnetic Effect and Anomaly-Induced Transport,''
  Prog.\ Part.\ Nucl.\ Phys.\  {\bf 75} (2014) 133
  [arXiv:1312.3348 [hep-ph]].

\bibitem{Landsteiner:2016led}
  K.~Landsteiner,
  ``Notes on Anomaly Induced Transport,''
  Acta Phys.\ Polon.\ B {\bf 47} (2016) 2617
  [arXiv:1610.04413 [hep-th]].

\bibitem{Landsteiner:2011cp}
  K.~Landsteiner, E.~Megias and F.~Pena-Benitez,
  ``Gravitational Anomaly and Transport,''
  Phys.\ Rev.\ Lett.\  {\bf 107} (2011) 021601
  [arXiv:1103.5006 [hep-ph]].
  
\bibitem{Landsteiner:2011iq}
  K.~Landsteiner, E.~Megias, L.~Melgar and F.~Pena-Benitez,
  ``Holographic Gravitational Anomaly and Chiral Vortical Effect,''
  JHEP {\bf 1109} (2011) 121
  [arXiv:1107.0368 [hep-th]].

\bibitem{Jensen:2012kj}
  K.~Jensen, R.~Loganayagam and A.~Yarom,
  JHEP {\bf 1302} (2013) 088
  [arXiv:1207.5824 [hep-th]].

 
\bibitem{Jensen:2013rga}
  K.~Jensen, R.~Loganayagam and A.~Yarom,
  ``Chern-Simons terms from thermal circles and anomalies,''
  JHEP {\bf 1405} (2014) 110
  [arXiv:1311.2935 [hep-th]].

\bibitem{Stone:2018zel}
  M.~Stone and J.~Kim,
  ``Mixed Anomalies: Chiral Vortical Effect and the Sommerfeld Expansion,''
  Phys. Rev. D {\bf 98}, 025012 (2018) 
  [arXiv:1804.08668 [cond-mat.mes-hall]].

  
\bibitem{Golkar:2015oxw}
  S.~Golkar and S.~Sethi,
  ``Global Anomalies and Effective Field Theory,''
  JHEP {\bf 1605} (2016) 105
  [arXiv:1512.02607 [hep-th]].

\bibitem{Glorioso:2017lcn} 
  P.~Glorioso, H.~Liu and S.~Rajagopal,
  ``Global Anomalies, Discrete Symmetries, and Hydrodynamic Effective Actions,''
  arXiv:1710.03768 [hep-th].

\bibitem{Vilenkin:1980zv} 
  A.~Vilenkin,
  ``Quantum Field Theory At Finite Temperature In A Rotating System,''
  Phys.\ Rev.\ D {\bf 21}, 2260 (1980).

\bibitem{Ambrus:2014uqa} 
  V.~E.~Ambru\c{s} and E.~Winstanley,
  ``Rotating quantum states,''
  Phys.\ Lett.\ B {\bf 734}, 296 (2014)
  [arXiv:1401.6388 [hep-th]].

\bibitem{Ambrus:2015lfr}
  V.~E.~Ambrus and E.~Winstanley,
  ``Rotating fermions inside a cylindrical boundary,''
  Phys.\ Rev.\ D {\bf 93} (2016) no.10,  104014
  [arXiv:1512.05239 [hep-th]].


\bibitem{Chernodub:2016kxh}
  M.~N.~Chernodub and S.~Gongyo,
  ``Interacting fermions in rotation: chiral symmetry restoration, moment of inertia and thermodynamics,''
  JHEP {\bf 1701} (2017) 136
  [arXiv:1611.02598 [hep-th]].

\bibitem{Chernodub:2017ref} 
  M.~N.~Chernodub and S.~Gongyo,
``Effects of rotation and boundaries on chiral symmetry breaking of relativistic fermions,''
  Phys.\ Rev.\ D {\bf 95}, no. 9, 096006 (2017)
  [arXiv:1702.08266 [hep-th]].

\bibitem{Chernodub:2017mvp} 
  M.~N.~Chernodub and S.~Gongyo,
 ``Edge states and thermodynamics of rotating relativistic fermions under magnetic field,''
  Phys.\ Rev.\ D {\bf 96}, no. 9, 096014 (2017)
  [arXiv:1706.08448 [hep-th]].
  

\bibitem{Ebihara:2016fwa}
  S.~Ebihara, K.~Fukushima and K.~Mameda,
 ``Boundary effects and gapped dispersion in rotating fermionic matter,''
  Phys.\ Lett.\ B {\bf 764} (2017) 94
  [arXiv:1608.00336 [hep-ph]].
  
\bibitem{Avkhadiev:2017fxj}
  A.~Avkhadiev and A.~V.~Sadofyev,
  ``Chiral Vortical Effect for Bosons,''
  Phys.\ Rev.\ D {\bf 96} (2017) no.4,  045015
  [arXiv:1702.07340 [hep-th]].

\bibitem{Yamamoto:2017uul}
  N.~Yamamoto,
  ``Photonic chiral vortical effect,''
  Phys.\ Rev.\ D {\bf 96} (2017) no.5,  051902
  [arXiv:1702.08886 [hep-th]].
  
\bibitem{Zyuzin}
 V.~A.~ Zyuzin,
  ``Landau levels for electromagnetic wave,''
  Phys.\ Rev.\ A {\bf 96}, 043830 (2017)
  [arXiv:1610.08048v2 [physics.optics]]

\bibitem{Dolgov:1988qx}
  A.~D.~Dolgov, I.~B.~Khriplovich, A.~I.~Vainshtein and V.~I.~Zakharov,
  ``Photonic Chiral Current and Its Anomaly in a Gravitational Field,''
  Nucl.\ Phys.\ B {\bf 315} (1989) 138.
 
\bibitem{Agullo:2016lkj}
  I.~Agullo, A.~del Rio and J.~Navarro-Salas,
  ``Electromagnetic duality anomaly in curved spacetimes,''
  Phys.\ Rev.\ Lett.\  {\bf 118} (2017) no.11,  111301
  [arXiv:1607.08879 [gr-qc]].

\bibitem{ref:Calkin}
   M.~G.~Calkin, 
   ``An invariance property of the free electromagnetic field,''
    Am.J.Phys. {\bf 33}, 958 (1965).

\bibitem{Lipkin}
	H.~Lipkin,
	``Existence of a New Conservation Law in Electromagnetic Theory,''
	Journal of Mathematical Physics 5, 696 (1964).
	
\bibitem{Kibble}
	T.W.B.~Kibble,
	``Conservation Laws for Free Fields,''
	Journal of Mathematical Physics 6, 1022 (1965).

	
\bibitem{Deser:1976iy}
  S.~Deser and C.~Teitelboim,
  ``Duality Transformations of Abelian and Nonabelian Gauge Fields,''
  Phys.\ Rev.\ D {\bf 13} (1976) 1592.
 
\bibitem{tangcohen}
	Y.~Tang and A.E.~Cohen,
	``Optical Chirality and Its Interaction with Matter,''
	Phys. Rev. Lett. 104, 163901 (2010).

\bibitem{Bliokh:2014pva} 
  K.~Y.~Bliokh, Y.~S.~Kivshar and F.~Nori,
  ``Magnetoelectric Effects in Local Light-Matter Interactions,''
  Phys.\ Rev.\ Lett.\  {\bf 113}, no. 3, 033601 (2014)
  [arXiv:1312.4325 [physics.optics]].

\bibitem{Bliokh:2018ehc} 
  F.~Alpeggiani, K.~Y.~Bliokh, F.~Nori and L. Kuipers
  ``Electromagnetic helicity in complex media''
  Phys. Rev. Lett. {\bf 120}, 243605 (2018)
  [arXiv:1802.09392 [physics.optics]].
  

\bibitem{Elbistan:2016xzq} 
  M.~Elbistan, P.~A.~Horvathy and P.-M.~Zhang,
  ``Duality and helicity: the photon wave function approach,''
  Phys.\ Lett.\ A {\bf 381}, 2375 (2017)
  [arXiv:1608.08573 [hep-th]].
  
\bibitem{Elbistan:2018fkr} 
  M.~Elbistan,
  ``Optical helicity and Hertz vectors,''
  Phys.\ Lett.\ A {\bf 382}, 1897 (2018)
  [arXiv:1802.10485 [physics.optics]].

\bibitem{ref:Cameron}
  R.~P.~Cameron, S.~M.~Barnett and A.~M.~Yao,
 ``Optical helicity, optical spin and related quantities in electromagnetic theory,'' 
  New J.Phys. {\bf 14}, 053050 (2012).

\bibitem{Iyer:1982ah} 
  B.~R.~Iyer,
  ``Dirac Field Theory In Rotating Coordinates,''
  Phys.\ Rev.\ D {\bf 26}, 1900 (1982).

\bibitem{ref:pathologies:1}
 O. Levin, Y. Peleg and A. Peres,
  ``Unruh effect for circular motion in a cavity,'' 
  J.Phys. {\bf A26}, 3001 (1993);
  V.~A.~De Lorenci and N.~F.~Svaiter,
  ``A Rotating quantum vacuum,''
  Found.\ Phys.\  {\bf 29}, 1233 (1999).
  
\bibitem{ref:pathologies:2}
  P.~C.~W.~Davies, T.~Dray and C.~A.~Manogue,
  ``The Rotating quantum vacuum,''
  Phys.\ Rev.\ D {\bf 53}, 4382 (1996).

\end{thebibliography}
\end{document}